\documentclass[aps,pra,10pt,twocolumn,nofootinbib]{revtex4-2}%
\pdfoutput=1%
\usepackage{graphicx}%
\usepackage{amsmath, amssymb}%
\usepackage{mathphyscmd}%
\everymath{\displaystyle}%

\usepackage{hyperref} %
\hypersetup{
	pdfinfo={%
		Author       = {David Gaspard},
		Title        = {Transverse quantum decoherence of a fast particle in a gas},
		Subject      = {Decoherence theory},
		CreationDate = {D:20220823125039+0200}, %
	},
	pdfborderstyle = {/S/U/W 1}, %
	breaklinks = true,
}

\begin{document}%
\title{Transverse quantum decoherence of a fast particle in a gas}%

\author{David Gaspard}%
\author{Jean-Marc Sparenberg}%
\affiliation{Nuclear Physics and Quantum Physics, CP229, Universit\'e libre de Bruxelles (ULB), B-1050 Brussels, Belgium}%
\date{\today}%

\begin{abstract}%
The decoherence of a fast quantum particle in a gas is studied by applying the Kramers-Moyal expansion to the quantum master equation for the reduced density matrix of the particle.
This expansion leads to a general form of the Caldeira-Leggett master equation accounting for the angular variation of the differential cross section.
The equation describes the decoherence in both the longitudinal and transverse directions with respect to the particle motion.
It is shown that, when the differential cross section is concentrated in the forward direction, transverse decoherence dominates.
The coherence region off the diagonal of the density matrix is characterized by coherence lengths, which can be deduced, for Gaussian states, from the momentum covariance matrix according to a Heisenberg-type uncertainty relation.
Finally, the longitudinal-to-transverse ratio of the coherence lengths is estimated for an alpha particle of a few MeVs.
This ratio indicates that the coherence region looks like an ellipsoid elongated in the direction of motion.
\end{abstract}%
\maketitle%

\section{Introduction}\label{sec:introduction}
The question of the propagation of a quantum particle in a detector, such as a cloud chamber or an ionization chamber, is probably one of the least obvious ever addressed from first principles.
It was originally asked by Darwin~\cite{Darwin1929} and Mott~\cite{Mott1929} in the early years of quantum theory.
The problem at the time was to understand how the appearance of linear tracks of alpha particles in cloud chambers could be consistent with the quantum-mechanical description of these particles as waves.
In his paper, Mott gave a relevant explanation of this phenomenon by accounting for the whole Hilbert space generated by the excitation states of the detector constituents.
His explanation is still verified today in recent works~\cite{Cacciapuoti2007a, Cacciapuoti2007b, Figari2014, Carlone2015, DellAntonio2008, Sparenberg2018, GaspardD2019b}.
\par An important issue pointed out by Darwin and Mott resides in the lack of information about the state of the alpha particle provided by single-particle quantum measurements.
In particular, although the wave function of the alpha particle is depicted as symmetrical around the radioactive source, individual tracks seem to have preferred directions~\cite{Darwin1929, Mott1929}, in apparent contradiction with quantum mechanics.
This comes from the probabilistic nature of the quantum-mechanical wave function and the inherent difficulty of its interpretation.
However, the specific issue of interpretation is not the subject of the present paper.
Instead, this work attempts to describe on average the whole ensemble of possible measurement outcomes using the reduced density matrix of the particle.
It has been known since the pioneering work of Joos and Zeh~\cite{Zeh1970, Zeh1973, Joos1985} that the reduced density matrix of a particle undergoes decoherence, which is the exponential decay of its off-diagonal elements, due to the interaction with the quantum environment~\cite{Hornberger2003b, Hornberger2006a, Hornberger2007, Hornberger2009, Vacchini2009}.
This effect is considered as the signature of quantum measurement and the transition from quantum to classical behaviors~\cite{Joos1985, Gallis1990, Zurek1991, Zurek2003, Allahverdyan2013, Schlosshauer2019}.
Until now, decoherence theory has been successfully observed in numerous experiments on large slow molecules~\cite{Arndt1999, Hornberger2003a, Hackermuller2004, Juffmann2013, Eibenberger2013, Fein2019, Brand2020, Schrinski2020}, but not on fast incident particles.
\par The purpose of the present paper is to study the quantum master equation derived in the previous paper~\cite{GaspardD2022c} and to specialize it to the context of a fast particle passing through a detector modeled as a gas of slow particles.
As in the previous paper~\cite{GaspardD2022c}, the interaction potential between the incident particle and the gas particles is assumed to be short ranged, hence neglecting the Coulombic nature of the collisions with the electrons~\cite{Segre1953, Segre1977, Ziegler1985, Sigmund2006, Sigmund2014}.
Despite this approximation, the differential cross section in the center-of-mass frame is supposed to be focused enough in the forward direction to avoid significant deviation of the particle from its initial direction~\cite{Wax1998, Cheng1999}.
This latter assumption enables the use of a quantum-mechanical analog of the Kramers-Moyal expansion~\cite{Kramers1940, Moyal1949b, Akama1965, Pawula1967} to approximate the collisional term of the master equation.
This procedure leads to an equation similar to the master equation found by Caldeira and Leggett~\cite{Caldeira1983b, Diosi1993b, Breuer2002, Vacchini2009, Hornberger2009, Kamleitner2010} with drift and diffusion terms acting in momentum space, but of a more general kind, because it does not assume that the particle is slow compared to the gas.
In the Wigner representation, this master equation reduces to a Fokker-Planck equation for the momentum distribution of the particle.
Using a definition for the coherence length~\cite{Barnett2000}, it will be shown that, due to the assumption of small deflection of the particle, decoherence dominates in the direction transverse to motion.
Therefore, the coherent wave packet looks like an ellipsoid elongated in the direction of motion.
\par This paper is organized as follows.
The quantum master equation of Ref.~\cite{GaspardD2022c} governing the reduced density matrix of the particle is presented in Sec.~\ref{sec:general-properties}.
The general properties of this equation, including thermalization and spatial decoherence, are considered in Secs.~\ref{sec:energy-transfer} and~\ref{sec:decoherence-rate}, respectively.
These general properties do not rely on the Kramers-Moyal approximation.
Then, approximate expressions based on the Kramers-Moyal expansion are derived in Sec.~\ref{sec:kramers-moyal}.
In particular, after the calculation of the moments that takes place in Sec.~\ref{sec:moments}, a general form of the Caldeira-Leggett master equation is presented in Sec.~\ref{sec:caldeira-leggett}.
This equation is then specialized to the case of strongly forward scattering in Sec.~\ref{sec:forward-scattering} and expressed in terms of the angular momentum operator to shed light on the spherical nature of transverse decoherence.
Additional explanations about the appearance of the angular momentum operator are given in Appendix~\ref{app:decoherence-on-sphere}. 
The Wigner transform of the master equation is then established in Sec.~\ref{sec:wigner-representation}.
In Sec.~\ref{sec:coherence-length}, a coherence length matrix is introduced to characterize the region of coherence around the diagonal of the reduced density matrix.
It is shown in Sec.~\ref{sec:momentum-variance} that, for the Gaussian states predicted by the Fokker-Planck equation, the coherence length is the inverse of the standard deviation of the momentum.
This is due to the existence of a Heisenberg-type uncertainty relation between the coherence length and the momentum.
Finally, this relation leads to an approximate solution for the time evolution of the coherence length which is obtained in Sec.~\ref{sec:coherence-evolution} based on the calculations of Sec.~\ref{sec:momentum-evolution}.
\par Throughout this paper, SI units are used with the recommended values of Ref.~\cite{Tiesinga2021} for the fundamental constants.
Regarding the notations, $c$ is the speed of light in vacuum, $h$ is the Planck constant, $\hbar=h/2\pi$ is the reduced Planck constant, $\alpha$ is the fine-structure constant, and $\kbol$ is the Boltzmann constant.
All the calculations will be performed for an arbitrary number of spatial dimensions: $d\in\{2,3,4,\ldots\}$.
Quantum operators will be denoted by $\op{A}$ to distinguish them from the associated eigenvalue $A$.

\section{General properties of the master equation}\label{sec:general-properties}
One considers a model for a spinless quantum particle of mass $m_\sys$ interacting with a thermal bath of $N$ mobile scatterers of mass $m_\bath$ in a cubic enclosure denoted by $\mathcal{V}$ of side $L$ and volume $V=L^d$.
The Hamiltonian of the whole system reads~\cite{GaspardD2022c}
\begin{equation}\label{eq:general-hamiltonian}
\op{H} = \underbrace{\frac{\hbar^2\op{\vect{k}}_\sys^2}{2m_\sys}}_{\op{H}_\sys} + \underbrace{\sum_{i=1}^N \frac{\hbar^2\op{\vect{k}}_i^2}{2m_\bath}}_{\op{H}_\bath} + \underbrace{\sum_{i=1}^N u(\op{\vect{r}} - \op{\vect{x}}_i)}_{\op{U}}  \:,
\end{equation}
where $(\op{\vect{r}},\op{\vect{k}}_\sys)$ are the position and the momentum of the particle, and $(\op{\vect{x}}_1,\op{\vect{x}}_2,\ldots,\op{\vect{x}}_N)$ and $(\op{\vect{k}}_1,\op{\vect{k}}_2,\ldots,\op{\vect{k}}_N)$ are the positions and the momenta of the scatterers, respectively.
These momenta are quantized to the cubic lattice
\begin{equation}\label{eq:quantized-momentum}
\vect{k} = \frac{2\pi}{L}\tran{(n_1,\ldots,n_d)}  \quad\forall (n_1,\ldots,n_d)\in\mathbb{Z}^d \:,
\end{equation}
due to the periodic boundary conditions over the wave function.
The number density of scatterers, $n=N/V$, is supposed to be fixed, even in the continuum limit for the momentum states, namely $V\rightarrow\infty$.
In addition, the potential $u(\vect{r})$ is assumed to be spherically symmetric and of short range.
\par It was shown in the previous paper~\cite{GaspardD2022c} that the master equation for the reduced density matrix of the particle, defined as the partial trace over the bath of the full density matrix
\begin{equation}\label{eq:def-reduced-density-matrix}
\op{\rho}_\sys(t) = \Tr_\bath \op{\rho}(t)  \:,
\end{equation}
has the form
\begin{equation}\label{eq:simplified-redfield}
\pder{\op{\rho}_\sys}{t} = \supop{L}_\sys\op{\rho}_\sys + \supop{R}_\sys\op{\rho}_\sys  \:.
\end{equation}
The notation $\supop{L}_\sys$ in Eq.~\eqref{eq:simplified-redfield} stands for the Liouvillian superoperator
\begin{equation}\label{eq:liouvillian-supop}
\supop{L}_\sys\op{\rho}_\sys = \frac{1}{\I\hbar} [\op{H}_\sys, \op{\rho}_\sys]  \:,
\end{equation}
which describes the free propagation of the particle, and $\supop{R}_\sys$ is the Redfieldian superoperator, or dissipator~\cite{Breuer2002}, which reads~\cite{GaspardD2022c}
\begin{equation}\label{eq:redfieldian-supop}
\supop{R}_\sys\op{\rho}_\sys = \frac{1}{2} \sum_{\vect{q}} \left( \E^{\I\vect{q}\cdot\op{\vect{r}}} \{\op{W}_{\vect{q}}, \op{\rho}_\sys\} \E^{-\I\vect{q}\cdot\op{\vect{r}}} - \{\op{W}_{\vect{q}}, \op{\rho}_\sys\} \right)  \:,
\end{equation}
where $\{\op{A},\op{B}\}=\op{A}\op{B}+\op{B}\op{A}$ is the anticommutator.
The dissipator~\eqref{eq:redfieldian-supop} was derived in the previous paper~\cite{GaspardD2022c} using the Markov assumption, the perturbative approximation, and the weak scattering condition~\cite{ShengP2006, Akkermans2007}
\begin{equation}\label{eq:weak-scattering}
k_{\sys,0}\lscat \gg 1  \:,
\end{equation}
where $k_{\sys,0}=2\pi/\lambda_{\sys,0}$ is the initial particle wave number and $\lscat=(n\sigma)^{-1}$ is the mean free path, $\sigma$ being the total collision cross section.
It was shown in that paper that the Redfield equation~\eqref{eq:simplified-redfield}--\eqref{eq:redfieldian-supop} approximately reduces to the Boltzmann equation in the Wigner representation.
In this respect, it is similar to the quantum linear Boltzmann equation of Refs.~\cite{Hornberger2006a, Hornberger2007, Vacchini2009} although it cannot be cast into the Lindblad form.
\par In Eq.~\eqref{eq:redfieldian-supop}, $\op{W}_{\vect{q}}$ denotes the total collision rate operator associated with the transferred momentum $\vect{q}$, and given by
\begin{equation}\label{eq:def-rate-operator}
\op{W}_{\vect{q}} = W_{\vect{q}}(\op{\vect{k}}_\sys) = N \avg{ w_{\vect{q}}(\op{\vect{k}}_\sys, \op{\vect{k}}_\bath) }_\bath  \:,
\end{equation}
where the notation $\tavg{\cdot}_\bath$ stands for the average over the momentum $\vect{k}_\bath$ of some generic bath particle.
This average is defined as
\begin{equation}\label{eq:def-bath-avg}
\avg{X(\op{\vect{k}}_\sys, \op{\vect{k}}_\bath)}_\bath = \int_{\mathbb{R}^d} \D\vect{k}_\bath f_\bath(\vect{k}_\bath) X(\op{\vect{k}}_\sys, \vect{k}_\bath)  \:,
\end{equation}
and is still an operator acting on particle ``$\sys$''.
In Eq.~\eqref{eq:def-bath-avg}, $f_\bath(\vect{k}_\bath)$ is the single-particle momentum distribution of the bath normalized according to $\textstyle\int_{\mathbb{R}^d}\D\vect{k}_\bath f_\bath(\vect{k}_\bath)=1$.
The gas is supposed to be in the rest frame so that the average velocity is zero: $\tavg{\vect{k}_\bath}_\bath=\vect{0}$.
\par
\begin{figure}[ht]%
\includegraphics{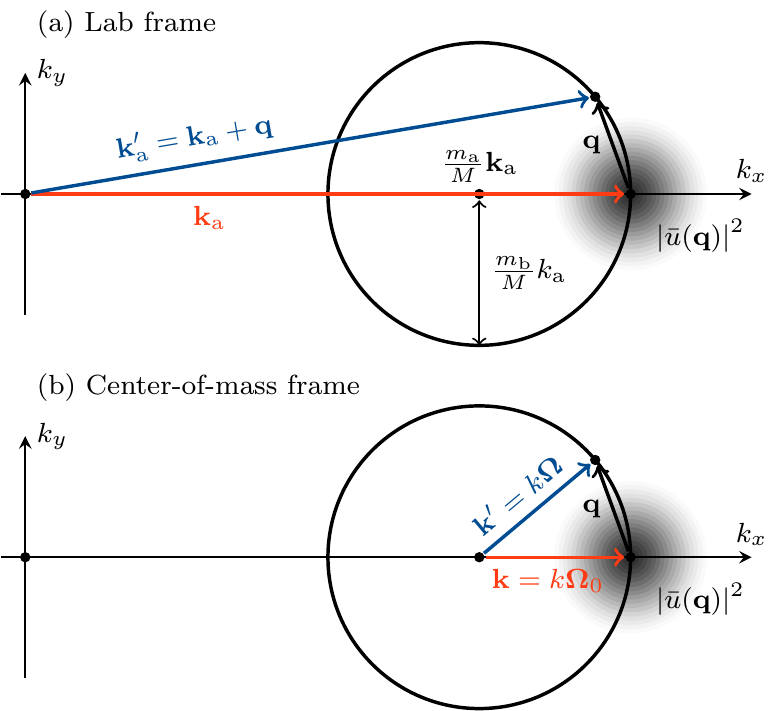}%
\caption{Distribution of the final momentum $\vect{k}'_\sys=\vect{k}_\sys+\vect{q}$ of particle ``$\sys$'' after a single collision according to Eqs.~\eqref{eq:def-rate-operator} and~\eqref{eq:def-binary-collision-rate}, assuming the target scatterer is initially at rest ($\vect{k}_\bath=\vect{0}$). The mass ratio is $\tfrac{m_\sys}{m_\bath}=3$.
The circle represents the spherical energy shell, $\delta(D_{\vect{q}})$, and the blurry function represents $\abs{\fourier{u}(\vect{q})}^2$ in Eq.~\eqref{eq:def-binary-collision-rate}.
Panel~(a): Situation in the lab frame. Panel~(b): Same situation in the center-of-mass frame described by the relative momenta $\vect{k}$ and $\vect{k}'$.}%
\label{fig:binary-collision-moments}%
\end{figure}

 In Eq.~\eqref{eq:def-rate-operator}, $w_{\vect{q}}(\op{\vect{k}}_\sys,\op{\vect{k}}_\bath)$ is the binary collision rate operator for a fixed value of $\op{\vect{k}}_\bath$.
At leading order of perturbation theory in the potential $u(\vect{r})$, this operator is given by
\begin{equation}\label{eq:def-binary-collision-rate}
w_{\vect{q}}(\op{\vect{k}}_\sys,\op{\vect{k}}_\bath) = \frac{2\pi}{\hbar} \frac{1}{V^2} \abs{\fourier{u}(\vect{q})}^2 \delta(\op{D}_{\vect{q}})  \:,
\end{equation}
with the energy difference operator defined as
\begin{equation}\label{eq:def-energy-diff-operator}
\op{D}_{\vect{q}} = E_{\op{\vect{k}}_\sys + \vect{q}} + E_{\op{\vect{k}}_\bath - \vect{q}} - E_{\op{\vect{k}}_\sys} - E_{\op{\vect{k}}_\bath}  \:.
\end{equation}
In Eq.~\eqref{eq:def-energy-diff-operator}, $E_{\op{\vect{k}}_\sys}=\tfrac{\hbar^2\op{\vect{k}}_\sys^2}{2m_\sys}$ and $E_{\op{\vect{k}}_\bath}=\tfrac{\hbar^2\op{\vect{k}}_\bath^2}{2m_\bath}$ are convenient notations for the free Hamiltonians of the particle and the generic bath scatterer, respectively.
The binary collision rate~\eqref{eq:def-binary-collision-rate} is schematically represented in Fig.~\ref{fig:binary-collision-moments} in the space of the transferred momentum $\vect{q}$ for a fixed value of $\vect{k}_\sys$ and for $\vect{k}_\bath=\vect{0}$.
The binary collision rate~\eqref{eq:def-binary-collision-rate} is the product of the energy conservation Dirac delta, $\delta(D_{\vect{q}})$, constraining $\vect{q}$ to a sphere, and the function $\abs{\fourier{u}(\vect{q})}^2$ centered at $\vect{q}=\vect{0}$.
\par In the continuum limit ($V\rightarrow\infty$), all the sums over the transferred momentum can be expressed in terms of the differential cross section $\textstyle\der{\sigma}{\Omega}$ according to the property~\cite{GaspardD2022c}
\begin{equation}\label{eq:rate-moment-property}\begin{split}
& \sum_{\vect{q}} W_{\vect{q}}(\vect{k}_\sys) F(\vect{q})  \\
& = \oint_{\mathcal{S}_d} \D\Omega \avg{nv\der{\sigma}{\Omega}(\vect{\Omega}\mid k\vect{\Omega}_0) F(k\vect{\Omega}-k\vect{\Omega}_0)}_\bath  \:,
\end{split}\end{equation}
which holds for any function $F(\vect{q})$ by definition of the differential cross section.
Property~\eqref{eq:rate-moment-property} will be often used throughout this paper.
In the right-hand side of Eq.~\eqref{eq:rate-moment-property}, $\vect{k}=k\vect{\Omega}_0$ is the relative momentum between the colliding particles defined as
\begin{equation}\label{eq:def-relative-momentum}
\vect{k} = \frac{m_\bath\vect{k}_\sys - m_\sys\vect{k}_\bath}{m_\sys + m_\bath}  \:,
\end{equation}
$\vect{v}=v\vect{\Omega}_0$ is the relative velocity
\begin{equation}\label{eq:def-relative-velocity}
\vect{v} = \vect{v}_\sys - \vect{v}_\bath = \frac{\hbar\vect{k}_\sys}{m_\sys} - \frac{\hbar\vect{k}_\bath}{m_\bath} = \frac{\hbar\vect{k}}{m}  \:,
\end{equation}
and $\vect{\Omega}_0$ and $\vect{\Omega}$ are respectively the initial and final directions of the relative motion between the particle and the bath scatterer, as shown in Fig.~\ref{fig:binary-collision-moments}(b).
They are both normalized to unity: $\norm{\vect{\Omega}_0}=\norm{\vect{\Omega}}=1$.
In Eq.~\eqref{eq:def-relative-velocity}, $m$ is the reduced mass of the binary system
\begin{equation}\label{eq:def-reduced-mass}
m = \frac{m_\sys m_\bath}{m_\sys + m_\bath}  \:.
\end{equation}
In addition, one also defines the total mass of the binary system for notational convenience
\begin{equation}\label{eq:def-total-mass}
M = m_\sys + m_\bath  \:.
\end{equation}
Finally, it should be noted that the average over the bath, $\tavg{\cdot}_\bath$, in Eq.~\eqref{eq:rate-moment-property}, concerns all the variables depending on $\vect{k}_\bath$, namely $v$, $k$, and $\vect{\Omega}_0$ in particular.

\subsection{Energy transfer and thermalization}\label{sec:energy-transfer}
In this section, it is shown that the Redfield equation~\eqref{eq:simplified-redfield} leads to the thermalization of the particle.
First, the governing equation for the mean particle energy, $\tavg{E_{\vect{k}_\sys}}_\sys=\Tr_\sys(E_{\op{\vect{k}}_\sys}\op{\rho}_\sys)$, is given by
\begin{equation}\label{eq:def-energy-evolution}
\der{\avg{E_{\vect{k}_\sys}}_\sys}{t} = \Tr_\sys\left(E_{\op{\vect{k}}_\sys} \supop{R}_\sys \op{\rho}_\sys\right)  \:,
\end{equation}
since the contribution from the free propagation term $\supop{L}_\sys\op{\rho}_\sys$ in Eq.~\eqref{eq:simplified-redfield} is zero due to energy conservation.
Using the cyclic property of trace and the momentum translation property
\begin{equation}\label{eq:translation-property}
\E^{-\I\vect{q}\cdot\op{\vect{r}}} f(\op{\vect{k}}_\sys) \E^{\I\vect{q}\cdot\op{\vect{r}}} = f(\op{\vect{k}}_\sys + \vect{q})  \:,
\end{equation}
one gets from Eqs.~\eqref{eq:redfieldian-supop} and~\eqref{eq:def-energy-evolution}
\begin{equation}\label{eq:energy-evolution-1}
\der{\avg{E_{\vect{k}_\sys}}_\sys}{t} = \Tr_\sys\left( \frac{1}{2} \sum_{\vect{q}} \{\op{W}_{\vect{q}}, \op{\rho}_\sys\} (E_{\op{\vect{k}}_\sys+\vect{q}} - E_{\op{\vect{k}}_\sys}) \right)  \:.
\end{equation}
In addition, since the commutation $[\op{W}_{\vect{q}}, E_{\op{\vect{k}}_\sys}]=0$ holds, one can write
\begin{equation}\label{eq:energy-evolution-2}
\der{\avg{E_{\vect{k}_\sys}}_\sys}{t} = \Tr_\sys\left( \op{\rho}_\sys \sum_{\vect{q}} \op{W}_{\vect{q}} (E_{\op{\vect{k}}_\sys+\vect{q}} - E_{\op{\vect{k}}_\sys}) \right)  \:.
\end{equation}
Now, one can use property~\eqref{eq:rate-moment-property} to deal with the sum over $\vect{q}$ in Eq.~\eqref{eq:energy-evolution-2}. Thus, one gets
\begin{equation}\label{eq:moment-energy-transfer}\begin{split}
& \sum_{\vect{q}} W_{\vect{q}}(\vect{k}_\sys) (E_{\vect{k}_\sys+\vect{q}} - E_{\vect{k}_\sys})  \\
& = \oint_{\mathcal{S}_d} \D\Omega \avg{nv\der{\sigma}{\Omega}(\vect{\Omega}\mid k\vect{\Omega}_0) (E_{\vect{k}_\sys + \vect{q}} - E_{\vect{k}_\sys})}_\bath  \:,
\end{split}\end{equation}
assuming implicitly $\vect{q}=k(\vect{\Omega}-\vect{\Omega}_0)$ on the right-hand side.
Introducing the energy transfer averaged over the differential cross section
\begin{equation}\label{eq:def-energy-transfer}
T_\sys = \oint_{\mathcal{S}_d} \D\Omega \der{\sigma}{\Omega}(\vect{\Omega}\mid k\vect{\Omega}_0) (E_{\vect{k}_\sys+\vect{q}} - E_{\vect{k}_\sys})  \:,
\end{equation}
Eq.~\eqref{eq:energy-evolution-2} can be rewritten more simply as
\begin{equation}\label{eq:energy-evolution-3}
\der{\avg{E_{\vect{k}_\sys}}_\sys}{t} = \Tr_\sys\left( \op{\rho}_\sys \avg{n\op{v} \op{T}_\sys}_\bath \right) = \avg{\avg{n\op{v} \op{T}_\sys}_\bath}_\sys  \:.
\end{equation}
Then, one expands the energy transfer integral~\eqref{eq:def-energy-transfer} to get
\begin{equation}\label{eq:energy-transfer-1}
T_\sys = \frac{\hbar^2}{2m_\sys} \oint_{\mathcal{S}_d} \D\Omega \der{\sigma}{\Omega}(\vect{\Omega}\mid k\vect{\Omega}_0) (\vect{q}^2 + 2\vect{k}_\sys\cdot\vect{q})  \:,
\end{equation}
To perform the integral~\eqref{eq:energy-transfer-1}, it is convenient to introduce the momentum-transfer cross section~\cite{Weinberg1958}
\begin{equation}\label{eq:def-transfer-cross-section}
\sigma_{\rm tr}(k) = \oint_{\mathcal{S}_d} \D\Omega \der{\sigma}{\Omega}(\vect{\Omega}\mid k\vect{\Omega}_0) \left(1 - \vect{\Omega}\cdot\vect{\Omega}_0\right)  \:.
\end{equation}
This quantity has the same units as the total cross section $\sigma(k)$, and is always bounded by $\sigma_{\rm tr}\in[0,\sigma]$.
It mainly characterizes the dispersion of the center-of-mass differential cross section around the forward direction.
When $\sigma_{\rm tr}\ll\sigma$, this means that the cross section is peaked in the forward direction and that the particle undergoes small deflections upon collision.
On the contrary, when $\sigma_{\rm tr}=\sigma$, the cross section is independent of the direction, i.e., isotropic, and the particle is strongly deflected.
\par Using the facts that $\vect{q}=k(\vect{\Omega}-\vect{\Omega}_0)$ and $\vect{k}=k\vect{\Omega}_0$, one finds two exact integrals for the transferred momentum which derive from definition~\eqref{eq:def-transfer-cross-section}, namely
\begin{equation}\label{eq:momentum-transfer-1}
\oint_{\mathcal{S}_d} \D\Omega \der{\sigma}{\Omega}(\vect{\Omega}\mid k\vect{\Omega}_0) \vect{q} = -\vect{k}\sigma_{\rm tr}(k)  \:,
\end{equation}
and
\begin{equation}\label{eq:momentum-transfer-2}
\oint_{\mathcal{S}_d} \D\Omega \der{\sigma}{\Omega}(\vect{\Omega}\mid k\vect{\Omega}_0) \vect{q}^2 = 2\vect{k}^2\sigma_{\rm tr}(k)  \:.
\end{equation}
Substituting Eqs.~\eqref{eq:momentum-transfer-1} and~\eqref{eq:momentum-transfer-2} into Eq.~\eqref{eq:energy-transfer-1}, one gets
\begin{equation}\label{eq:energy-transfer-2}
T_\sys = \sigma_{\rm tr}(k) \frac{\hbar^2}{m_\sys} (\vect{k}^2 - \vect{k}_\sys\cdot\vect{k})  \:.
\end{equation}
Using Eq.~\eqref{eq:def-relative-momentum} to expand $\vect{k}$, Eq.~\eqref{eq:energy-transfer-2} becomes
\begin{equation}\label{eq:energy-transfer-3}
T_\sys = \sigma_{\rm tr}(k) \frac{\hbar^2}{M^2} \left(m_\sys\vect{k}_\bath^2 - m_\bath\vect{k}_\sys^2 + (m_\sys-m_\bath)\vect{k}_\sys\cdot\vect{k}_\bath\right)  \:.
\end{equation}
To evaluate the averages in Eq.~\eqref{eq:energy-evolution-3}, one considers several approximations.
First, one assumes that
\begin{equation}\label{eq:uncorrelated-momentum-approx}
\avg{v\sigma_{\rm tr}(k)\vect{k}_\sys\cdot\vect{k}_\bath}_\bath \simeq 0  \:.
\end{equation}
This is reasonable since the particle and bath momenta are uncorrelated and the gas is supposed to be at rest: $\tavg{\vect{k}_\bath}_\bath=\vect{0}$.
In the same spirit, one assumes that $v\sigma_{\rm tr}(k)$ varies slowly enough around the mass center of the momentum distributions $\rho_\sys(\vect{k}_\sys)$ and $\rho_\bath(\vect{k}_\bath)$ to use the following approximations
\begin{equation}\label{eq:momentum-average-approx-1}\begin{cases}
\avg{v\sigma_{\rm tr}(k)\vect{k}_\bath^2}_\bath \simeq \avg{v\sigma_{\rm tr}(k)}_\bath \avg{\vect{k}_\bath^2}_\bath  \:,\\
\avg{v\sigma_{\rm tr}(k)\vect{k}_\sys^2}_\sys   \simeq \avg{v\sigma_{\rm tr}(k)}_\sys \avg{\vect{k}_\sys^2}_\sys     \:.
\end{cases}\end{equation}
From Eqs.~\eqref{eq:energy-evolution-3} and~\eqref{eq:energy-transfer-3}, a simple closed equation for the mean particle energy is finally obtained
\begin{equation}\label{eq:energy-evolution-final}
\der{\avg{E_{\vect{k}_\sys}}}{t} = 2\eta \left(\avg{E_{\vect{k}_\bath}} - \avg{E_{\vect{k}_\sys}}\right)  \:,
\end{equation}
where $\eta$ is an energy-based friction coefficient defined as
\begin{equation}\label{eq:def-energy-friction-param}
\eta = \frac{m_\sys m_\bath}{M^2} \avg{nv\sigma_{\rm tr}(k)}  \:.
\end{equation}
In Eqs.~\eqref{eq:energy-evolution-final} and~\eqref{eq:def-energy-friction-param}, $\tavg{\cdot}$ denotes the average over the momenta of both the bath and the particle.
\par If $\eta$ is assumed to be time independent, then the solution of Eq.~\eqref{eq:energy-evolution-final} simply reads
\begin{equation}\label{eq:energy-thermalization}
\avg{E_{\vect{k}_\sys}(t)} = \left(\avg{E_{\vect{k}_\sys}(0)} - \avg{E_{\vect{k}_\bath}}\right)\E^{-2\eta t} + \avg{E_{\vect{k}_\bath}}  \:,
\end{equation}
for any initial energy $\tavg{E_{\vect{k}_\sys}(0)}$ for the particle.
The solution~\eqref{eq:energy-thermalization} implies that the mean particle energy always tends to the mean energy $\tavg{E_{\vect{k}_\bath}}=\tfrac{d}{2}\kbol T$ of the generic bath scatterer, $T$ being the absolute temperature.
This result is consistent with the thermalization process whereby the quantum particle reaches thermal equilibrium with the bath.
\par In the framework of swift charged particles traveling through matter, the governing equation of the mean energy is the famous Bethe formula~\cite{Segre1953, Segre1977, Ziegler1985, Sigmund2006, Sigmund2014}
\begin{equation}\label{eq:bethe-formula-general}
\der{\avg{E_{\vect{k}_\sys}}}{x} = -S(\avg{E_{\vect{k}_\sys}})  \:,
\end{equation}
where $S$ denotes the stopping power, which is a known function of the energy, and $x$ is the total path length of the particle.
The stopping power mainly accounts for the Coulomb interaction between the swift particle and the electrons of the medium.
If the contribution from the thermal bath is neglected in Eq.~\eqref{eq:energy-evolution-final}, then Eqs.~\eqref{eq:energy-evolution-final} and~\eqref{eq:bethe-formula-general} are completely analogous, and the energy-dependent friction parameter $\eta$ can be identified as
\begin{equation}\label{eq:energy-friction-from-stopping}
\eta = \frac{\sqrt{\avg{\vect{v}_\sys^2}}}{2\avg{E_{\vect{k}_\sys}}} S(\avg{E_{\vect{k}_\sys}})  \:,
\end{equation}
since the path length element is on average given by $\D x^2=\tavg{\vect{v}_\sys^2}\D t^2$.
In principle, Eq.~\eqref{eq:energy-friction-from-stopping} can be used to adjust the value of $\eta$ on experimental measurements of the stopping power~\cite{ESTAR}.
It is remarkable that such an adjustment is feasible, despite the fact that long-range interactions, such as the Coulomb interaction, are beyond the scope of validity of the present model.
\par The same reasoning based on Eq.~\eqref{eq:def-energy-evolution} can be applied to the mean momentum $\tavg{\vect{k}_\sys}$.
The calculations are very similar to the previous ones, and the result reads
\begin{equation}\label{eq:momentum-evolution-final}
\der{\avg{\vect{k}_\sys}}{t} = -\zeta \avg{\vect{k}_\sys}  \:,
\end{equation}
where $\zeta$ is a momentum-based friction coefficient defined as
\begin{equation}\label{eq:def-momentum-friction-param}
\zeta = \frac{m_\bath}{M} \avg{nv\sigma_{\rm tr}(k)}  \:.
\end{equation}
Equation~\eqref{eq:momentum-evolution-final} shows that, in contrast to the mean energy, the mean momentum tends to zero upon thermalization.
Note, in addition, that the characteristic relaxation rate denoted by $\zeta$ in Eq.~\eqref{eq:momentum-evolution-final} is different from $\eta$.
The momentum-based parameter is always larger ($\zeta>\eta$), although this difference is negligible for a very heavy particle ($m_\sys\gg m_\bath$).
\par Finally, in the long-time limit ($t\rightarrow\infty$), the reduced density matrix of the particle tends to the equilibrium Boltzmann distribution
\begin{equation}\label{eq:equilibrium-distrib}
\op{\rho}_\sys \propto \E^{-\hbar^2\op{\vect{k}}_\sys^2/(2m_\sys\kbol T)}  \:.
\end{equation}
This is typically shown by applying the detailed balance condition~\cite{Breuer2002} to Eq.~\eqref{eq:redfieldian-supop}.
It can also be interpreted as a consequence of the relaxation process described by the Fokker-Planck equation derived in Sec.~\ref{sec:kramers-moyal}.

\subsection{Decoherence in position space}\label{sec:decoherence-rate}
In this section, one considers the other important property of the Redfield equation~\eqref{eq:simplified-redfield}, namely the decoherence.
In particular, one seeks to know how decoherence affects the density matrix in the position space.
To this end, one assumes that the density matrix of the particle is initially in the momentum eigenstate at $t=0$
\begin{equation}\label{eq:initial-pure-state}
\op{\rho}_\sys(0) = \ket{\vect{k}_{\sys,0}}\bra{\vect{k}_{\sys,0}}  \:.
\end{equation}
In this particular case, $\op{\rho}_\sys(t)$ remains diagonal in momentum space due to the gas uniformity, so that one has $[\op{H}_\sys,\op{\rho}_\sys]=0~\forall t>0$, and only the dissipator survives in Eq.~\eqref{eq:simplified-redfield}.
Under such circumstances, one finds in position representation:
\begin{equation}\label{eq:redfield-pos-basis-1}
\pder{\rho_\sys}{t}(\vect{r},\dual{\vect{r}},t) = \bra{\vect{r}}\supop{R}_\sys\op{\rho}_\sys(t)\ket{\dual{\vect{r}}}  \:.
\end{equation}
Expanding the right-hand side of Eq.~\eqref{eq:redfield-pos-basis-1} by means of Eq.~\eqref{eq:redfieldian-supop} leads to
\begin{equation}\label{eq:redfield-pos-basis-2}
\pder{\rho_\sys}{t}(\vect{r},\dual{\vect{r}},t) = \sum_{\vect{q}} \left(\E^{\I\vect{q}\cdot(\vect{r}-\dual{\vect{r}})} - 1\right) W_{\vect{q}}(\vect{k}_{\sys,0}) \rho_\sys(\vect{r},\dual{\vect{r}},t)  \:,
\end{equation}
assuming that, at the beginning of the interaction with the gas, the momentum distribution of the incident particle remains close to the initial one in Eq.~\eqref{eq:initial-pure-state}.
Given the structure of Eq.~\eqref{eq:redfield-pos-basis-2}, it is useful to introduce what is often called the decoherence rate~\cite{Caldeira1983b, Joos1985, Gallis1990, Cheng1999, Hornberger2003b, Hornberger2009, Vacchini2009, Adler2006, Kamleitner2010, Schlosshauer2019, Breuer2002}
\begin{equation}\label{eq:def-complex-decoherence-rate}
F(\vect{s}) = \sum_{\vect{q}} \left(1 - \E^{\I\vect{q}\cdot\vect{s}}\right) W_{\vect{q}}(\vect{k}_{\sys,0})  \:,
\end{equation}
where $\vect{s}=\vect{r}-\dual{\vect{r}}$.
Equation~\eqref{eq:def-complex-decoherence-rate} means that the decoherence rate, $F(\vect{s})$, is directly given by the Fourier transform of the total collision rate $W_{\vect{q}}(\vect{k}_{\sys,0})$.
In the continuum limit ($V\rightarrow\infty$), Eq.~\eqref{eq:def-complex-decoherence-rate} can also be expressed in terms of the differential cross section according to Eq.~\eqref{eq:rate-moment-property}. One gets
\begin{equation}\label{eq:decoherence-rate-from-cross-section}
F(\vect{s}) = \oint_{\mathcal{S}_d} \D\Omega \avg{nv\der{\sigma}{\Omega}(\vect{\Omega}\mid k\vect{\Omega}_0) \left(1 - \E^{\I k(\vect{\Omega}-\vect{\Omega}_0)\cdot\vect{s}}\right)}_\bath  \:,
\end{equation}
where $\vect{k}=k\vect{\Omega}_0$ and $\vect{v}=v\vect{\Omega}_0$ are respectively given by Eqs.~\eqref{eq:def-relative-momentum} and~\eqref{eq:def-relative-velocity}, as usual, but for the special value $\vect{k}_\sys=\vect{k}_{\sys,0}$.
Using the notation $F(\vect{s})$, Eq.~\eqref{eq:redfield-pos-basis-2} reads
\begin{equation}\label{eq:redfield-pos-basis-3}
\pder{\rho_\sys}{t}(\vect{r},\dual{\vect{r}}) = -F(\vect{r}-\dual{\vect{r}}) \rho_\sys(\vect{r},\dual{\vect{r}})  \:,
\end{equation}
and can be easily integrated as follows:
\begin{equation}\label{eq:redfield-pos-basis-sol}
\rho_\sys(\vect{r},\dual{\vect{r}},t) = \E^{-F(\vect{r}-\dual{\vect{r}})t} \rho_\sys(\vect{r},\dual{\vect{r}},0)  \:.
\end{equation}
Since $F(\vect{s})$ is complex valued, it is convenient to separate its real and imaginary parts:
\begin{equation}\label{eq:decoherence-rate-imag-real}
F(\vect{s}) = \Re F(\vect{s}) + \I \Im F(\vect{s})  \:.
\end{equation}
The real part of $F(\vect{s})$ explicitly reads
\begin{equation}\label{eq:def-real-decoherence-rate}
\Re F(\vect{s}) = \sum_{\vect{q}} \left[ 1 - \cos(\vect{q}\cdot\vect{s}) \right] W_{\vect{q}}(\vect{k}_{\sys,0})  \:.
\end{equation}
Since $W_{\vect{q}}\geq 0$ and $\cos\alpha\leq 1$, this function is necessarily positive:
\begin{equation}\label{eq:real-decoherence-positivity}
\Re F(\vect{s}) \geq 0  \:.
\end{equation}
Therefore, according to the sign convention in Eqs.~\eqref{eq:redfield-pos-basis-3} and~\eqref{eq:redfield-pos-basis-sol}, the function $\Re F(\vect{s})$ represents the exponential decay rate of the off-diagonal entries of $\rho_\sys(\vect{r},\dual{\vect{r}})$.
This decay is the signature of decoherence in position space.
In contrast, the imaginary part of $F(\vect{s})$, given by
\begin{equation}\label{eq:def-imag-decoherence-rate}
\Im F(\vect{s}) = -\sum_{\vect{q}} \sin(\vect{q}\cdot\vect{s}) W_{\vect{q}}(\vect{k}_{\sys,0})  \:,
\end{equation}
represents the oscillation frequency of the off-diagonal entries.
At $\vect{r}=\dual{\vect{r}}$, the decoherence rate~\eqref{eq:def-complex-decoherence-rate} vanishes ($F(\vect{0})=0$).
Moreover, at large separation distance ($\norm{\vect{r}-\dual{\vect{r}}}\rightarrow\infty$), it reaches the saturation value~\cite{Gallis1990, Cheng1999, Hornberger2003b, Hornberger2009, Vacchini2009, Adler2006, Schlosshauer2019, Breuer2002}
\begin{equation}\label{eq:decoherence-saturation}
F(\vect{s}) \xrightarrow{\norm{\vect{s}}\rightarrow\infty} \sum_{\vect{q}} W_{\vect{q}}(\vect{k}_{\sys,0}) = W_{\rm tot}(\vect{k}_{\sys,0})  \:,
\end{equation}
which turns out to be equal to the total collision rate of the particle in the gas: $W_{\rm tot}(\vect{k}_{\sys,0})=\avg{nv\sigma(k)}_\bath$.
This saturated regime was successfully observed for matter waves in the 2000s~\cite{Hornberger2003a, Hackermuller2004, Hornberger2009}, and previously for light waves~\cite{Cheng1999}.

\section{Kramers-Moyal approximation}\label{sec:kramers-moyal}
In this section, approximations of the Redfield equation~\eqref{eq:simplified-redfield} are established based on the assumption that the deflections undergone by the particle due to the collisions are relatively small.
Under this assumption, it is appropriate to expand the first term of the dissipator~\eqref{eq:redfieldian-supop} using the Kramers-Moyal expansion~\cite{Kramers1940, Moyal1949b, Akama1965, Pawula1967}.
In the quantum-mechanical framework of Eq.~\eqref{eq:redfieldian-supop}, this expansion is achieved in practice by
\begin{equation}\label{eq:kramers-moyal-expansion}
\E^{\I\vect{q}\cdot\op{\vect{r}}} \op{X} \E^{-\I\vect{q}\cdot\op{\vect{r}}} \simeq \op{X} + \I q_i [\op{r}_i, \op{X}] + \frac{\I^2}{2!} q_i q_j [\op{r}_i, [\op{r}_j, \op{X}]]  \:,
\end{equation}
where $\op{X}$ stands for $\tfrac{1}{2}\{\op{W}_{\vect{q}}, \op{\rho}_\sys\}$.
In Eq.~\eqref{eq:kramers-moyal-expansion} and throughout this section, the implicit summation over the repeated indices $i,j\in\{1,2,\ldots,d\}$ will be used.
\par As one will see later, the assumption of small deflections is relevant in two different situations: either the incident particle is much heavier than the gas particles, that is
\begin{equation}\label{eq:kramers-moyal-condition-1}
m_\sys \gg m_\bath  \:,
\end{equation}
or the differential cross section $\textstyle\der{\sigma}{\Omega}$ is very peaked in the forward direction ($\vect{\Omega}=\vect{\Omega}_0$).
This latter condition can be formulated by requiring the transfer cross section $\sigma_{\rm tr}(k)$, defined in Eq.~\eqref{eq:def-transfer-cross-section}, to be much smaller than the total cross section $\sigma(k)$:
\begin{equation}\label{eq:kramers-moyal-condition-2}
\sigma_{\rm tr}(k) \ll \sigma(k)  \:.
\end{equation}
If at least one of the two conditions~\eqref{eq:kramers-moyal-condition-1} or~\eqref{eq:kramers-moyal-condition-2} is fulfilled, then the expansion~\eqref{eq:kramers-moyal-expansion} is justified.
Since they are not mutually exclusive, they can both be satisfied at the same time.
This will be the case, for instance, in Sec.~\ref{sec:coherence-evolution} when considering an alpha particle of a few MeVs.
\par Using Eq.~\eqref{eq:kramers-moyal-expansion} to expand the dissipator~\eqref{eq:redfieldian-supop}, one finds
\begin{equation}\label{eq:redfieldian-expansion-1}
\supop{R}_\sys\op{\rho}_\sys \simeq \frac{\I}{2} [\op{r}_i, \{\op{A}^{(1)}_i, \op{\rho}_\sys\}] - \frac{1}{4} [\op{r}_i, [\op{r}_j, \{\op{A}^{(2)}_{ij}, \op{\rho}_\sys\}]]  \:,
\end{equation}
where the tensor operators $\op{A}^{(1)}_i$ and $\op{A}^{(2)}_{ij}$ are respectively the first and second moments of the total collision rate $\op{W}_{\vect{q}}$.
They are defined as
\begin{equation}\label{eq:def-first-rate-moment}
\op{A}^{(1)}_i = \sum_{\vect{q}} q_i W_{\vect{q}}(\op{\vect{k}}_\sys)  \:,
\end{equation}
and
\begin{equation}\label{eq:def-second-rate-moment}
\op{A}^{(2)}_{ij} = \sum_{\vect{q}} q_i q_j W_{\vect{q}}(\op{\vect{k}}_\sys)  \:.
\end{equation}
The quantities~\eqref{eq:def-first-rate-moment} and~\eqref{eq:def-second-rate-moment} will be calculated in detail in Sec.~\ref{sec:moments}.
Finally, one assumes that the density matrix of the particle is approximately diagonal in the momentum basis
\begin{equation}\label{eq:diagonality-approx}
[\op{\vect{k}}_\sys, \op{\rho}_\sys] \simeq \vect{0}  \:.
\end{equation}
This will be the case for sufficiently broad wave packets.
As discussed in the previous paper~\cite{GaspardD2022c}, this condition also ensures the complete positivity of the Redfield equation~\eqref{eq:simplified-redfield}.
A consequence of Eq.~\eqref{eq:diagonality-approx} is that the anticommutators in Eq.~\eqref{eq:redfieldian-expansion-1} can be replaced by simple product
\begin{equation}\label{eq:redfieldian-expansion-2}
\supop{R}_\sys\op{\rho}_\sys \simeq \I [\op{r}_i, \op{A}^{(1)}_i\op{\rho}_\sys] - \frac{1}{2} [\op{r}_i, [\op{r}_j, \op{A}^{(2)}_{ij}\op{\rho}_\sys]]  \:.
\end{equation}
This approximation will be helpful later in Sec.~\ref{sec:caldeira-leggett} when introducing transverse decoherence.

\subsection{Calculation of the moments}\label{sec:moments}

\subsubsection{First moment}\label{sec:first-moment}
One calculates the first-order moment~\eqref{eq:def-first-rate-moment} in vector notations from the property~\eqref{eq:rate-moment-property}
\begin{equation}\label{eq:first-rate-moment-1}
\vect{A}^{(1)} = \oint_{\mathcal{S}_d} \D\Omega \avg{nv\der{\sigma}{\Omega}(\vect{\Omega}\mid k\vect{\Omega}_0) \vect{q}}_\bath  \:.
\end{equation}
Using Eq.~\eqref{eq:momentum-transfer-1}, one gets from Eq.~\eqref{eq:first-rate-moment-1}
\begin{equation}\label{eq:first-rate-moment-2}
\vect{A}^{(1)} = -\avg{nv\sigma_{\rm tr}(k)\vect{k}}_\bath  \:.
\end{equation}
This expression can be further simplified using an approximation similar to Eq.~\eqref{eq:uncorrelated-momentum-approx}:
\begin{equation}\label{eq:first-rate-moment-final}
\vect{A}^{(1)} = -\frac{m_\bath}{M} \avg{nv\sigma_{\rm tr}(k)}_\bath \vect{k}_\sys  \:.
\end{equation}
This is all for the first-order moment.
Note that the operator nature of $\vect{A}^{(1)}$ coming from the momentum $\op{\vect{k}}_\sys$ should be restored before substituting into Eq.~\eqref{eq:redfieldian-expansion-2}.

\subsubsection{Second moment}\label{sec:second-moment}
Next, one considers the second-order moment~\eqref{eq:def-second-rate-moment}.
Using again the property~\eqref{eq:rate-moment-property}, one can write in tensor notations
\begin{equation}\label{eq:second-rate-moment-1}
\matr{A}^{(2)} = \oint_{\mathcal{S}_d} \D\Omega \avg{nv\der{\sigma}{\Omega}(\vect{\Omega}\mid k\vect{\Omega}_0) \vect{q}\otimes\vect{q}}_\bath  \:,
\end{equation}
where ``$\otimes$'' denotes the dyadic product, or tensor product, of two vectors.
The quantity $\matr{A}^{(2)}$ is thus a $d\times d$ matrix.
To calculate the tensor in Eq.~\eqref{eq:second-rate-moment-1}, one decomposes the outgoing direction $\vect{\Omega}$ into the parallel and perpendicular components to the initial direction of motion $\vect{\Omega}_0$.
Denoting the unit vector in the transverse direction as $\vect{\Omega}_\perp$, one writes
\begin{equation}\label{eq:direction-decomposition}
\vect{\Omega} = \cos\theta \,\vect{\Omega}_0 + \sin\theta \,\vect{\Omega}_\perp  \:,
\end{equation}
with $\vect{\Omega}_0\cdot\vect{\Omega}_\perp=0$ by definition of $\vect{\Omega}_\perp$.
The integral over $\vect{\Omega}$ in Eq.~\eqref{eq:second-rate-moment-1} thus splits into two integrals over $\theta$ and $\vect{\Omega}_\perp$.
The differential element of solid angle in arbitrary dimension $d\geq 2$ reads~\cite{Olver2010}
\begin{equation}\label{eq:solid-angle-element}
\D\Omega = (\sin\theta)^{d-2} \D\theta \D\Omega_\perp  \:.
\end{equation}
Knowing that the cross section does not depend on the azimuthal direction, $\vect{\Omega}_\perp$, one gets
\begin{equation}\label{eq:second-rate-moment-4}
\matr{A}^{(2)} = \avg{ k^2nv S_{d-1} \int_0^{\pi}\D\theta (\sin\theta)^{d-2} \der{\sigma}{\Omega}(k,\theta) \matr{I}(\theta) }_\bath \:,
\end{equation}
using the angular tensor
\begin{equation}\label{eq:def-angular-tensor}
\matr{I}(\theta) = \oint_{\mathcal{S}_{d-1}} \frac{\D\Omega_\perp}{S_{d-1}} \left[ (\cos\theta-1)\vect{\Omega}_0 + \sin\theta\,\vect{\Omega}_\perp \right]^{\otimes 2}  \:,
\end{equation}
where $\vect{a}^{\otimes 2}=\vect{a}\otimes\vect{a}$ denotes the dyadic square, or tensor square.
The tensor~\eqref{eq:def-angular-tensor} can first be simplified due to the fact that the average of $\vect{\Omega}_\perp$ is zero.
This step removes the cross products of the form $\vect{\Omega}_0\otimes\vect{\Omega}_\perp$, and leads to
\begin{equation}\label{eq:angular-tensor-1}
\matr{I}(\theta) = \oint_{\mathcal{S}_{d-1}} \frac{\D\Omega_\perp}{S_{d-1}} (1-\cos\theta)^2 \vect{\Omega}_0^{\otimes 2} + \sin^2\theta\,\vect{\Omega}_\perp^{\otimes 2}  \:.
\end{equation}
The first term of Eq.~\eqref{eq:angular-tensor-1} is trivial, and the second term is given by
\begin{equation}\label{eq:angular-tensor-property}
\oint_{\mathcal{S}_{d-1}} \frac{\D\Omega_\perp}{S_{d-1}} \,\vect{\Omega}_\perp^{\otimes 2} = \frac{\matr{1} - \vect{\Omega}_0^{\otimes 2}}{d-1}  \:,
\end{equation}
where $\matr{1}$ represents the $d\times d$ identity matrix.
The quantity $(\matr{1} - \vect{\Omega}_0^{\otimes 2})$ on the right-hand side of Eq.~\eqref{eq:angular-tensor-property} is the projection matrix onto the plane transverse to the direction of motion.
Once diagonalized, this matrix is equal to $1$ in all the directions perpendicular to $\vect{\Omega}_0$, and $0$ in the parallel direction.
Thus, one arrives at the following expression of the angular tensor
\begin{equation}\label{eq:angular-tensor-result}
\matr{I}(\theta) = (1-\cos\theta)^2 \vect{\Omega}_0^{\otimes 2} + \sin^2\theta \frac{\matr{1} - \vect{\Omega}_0^{\otimes 2}}{d-1}  \:.
\end{equation}
When comparing to Eq.~\eqref{eq:second-rate-moment-4}, the tensor~\eqref{eq:angular-tensor-result} suggests to introduce two quadratic moments related to the cross section, namely
\begin{equation}\label{eq:cross-section-quad-moment-para}\begin{split}
\sigma_{\rm q\para}(k) & = \oint_{\mathcal{S}_d} \D\Omega \left(1 - \vect{\Omega}\cdot\vect{\Omega}_0\right)^2 \der{\sigma}{\Omega}(\vect{\Omega}\mid k\vect{\Omega}_0)  \\
 & = S_{d-1} \int_0^{\pi}\D\theta (\sin\theta)^{d-2} (1-\cos\theta)^2 \der{\sigma}{\Omega}(k,\theta)  \:,
\end{split}\end{equation}
for the longitudinal component, and
\begin{equation}\label{eq:cross-section-quad-moment-perp}\begin{split}
\sigma_{\rm q\perp}(k) & = \oint_{\mathcal{S}_d} \D\Omega \left(1 - (\vect{\Omega}\cdot\vect{\Omega}_0)^2\right) \der{\sigma}{\Omega}(\vect{\Omega}\mid k\vect{\Omega}_0) \\
 & = S_{d-1} \int_0^{\pi}\D\theta (\sin\theta)^{d-2} \sin^2\theta \der{\sigma}{\Omega}(k,\theta)  \:,
\end{split}\end{equation}
for the transverse component.
One notices that the arithmetic mean of these two quantities is exactly equal to the transfer cross section~\eqref{eq:def-transfer-cross-section}
\begin{equation}\label{eq:cross-section-quad-to-transfer}
\frac{\sigma_{\rm q\para}(k) + \sigma_{\rm q\perp}(k)}{2} = \sigma_{\rm tr}(k)  \:.
\end{equation}
Expression~\eqref{eq:cross-section-quad-to-transfer} will be useful to evaluate one of the two quantities when only the other is known.
Using the notations~\eqref{eq:cross-section-quad-moment-para} and~\eqref{eq:cross-section-quad-moment-perp}, Eq.~\eqref{eq:second-rate-moment-4} reduces to
\begin{equation}\label{eq:second-rate-moment-5}
\matr{A}^{(2)} = \avg{ k^2nv \left(\sigma_{\rm q\para}(k)\vect{\Omega}_0^{\otimes 2} + \sigma_{\rm q\perp}(k)\frac{\matr{1} - \vect{\Omega}_0^{\otimes 2}}{d-1}\right) }_\bath  \:.
\end{equation}
In addition, since the incident relative momentum is given by $\vect{k}=k\vect{\Omega}_0$, Eq.~\eqref{eq:second-rate-moment-5} can also be rewritten as
\begin{equation}\label{eq:second-rate-moment-6}
\matr{A}^{(2)} = \avg{ nv\sigma_{\rm q\para}(k)\vect{k}^{\otimes 2} + nv\sigma_{\rm q\perp}(k)\frac{k^2\matr{1} - \vect{k}^{\otimes 2}}{d-1} }_\bath  \:.
\end{equation}
In order to expand Eq.~\eqref{eq:second-rate-moment-6}, one assumes, as in Sec.~\ref{sec:energy-transfer}, that the cross section is a sufficiently smooth function of $k$ to use the approximation
\begin{equation}\label{eq:momentum-average-approx-2}
\avg{v\sigma_\mu(k) f(\vect{k})}_\bath \simeq \avg{v\sigma_\mu(k)}_\bath \avg{f(\vect{k})}_\bath  \:,
\end{equation}
where $\sigma_\mu(k)$ denotes any type of cross-sectional moments, including $\sigma_{\rm q\para}(k)$, $\sigma_{\rm q\perp}(k)$, or $\sigma_{\rm tr}(k)$, and $f(\vect{k})$ represents any function of $\vect{k}$.
On the one hand, the assumption~\eqref{eq:momentum-average-approx-2} allows to define collisional rates per unit time as
\begin{equation}\label{eq:def-collisional-rates}
\alpha_\mu(k_\sys) = \avg{nv\sigma_\mu(k)}_\bath  \:.
\end{equation}
On the other hand, it is possible to evaluate the average of $f(\vect{k})$ over the bath in this case.
Using Eq.~\eqref{eq:def-relative-momentum}, the average of $\vect{k}^{\otimes 2}$ reads
\begin{equation}\label{eq:momentum-dyadic-bath-avg}
\avg{\vect{k}^{\otimes 2}}_\bath = \frac{m_\bath^2}{M^2} \vect{k}_\sys^{\otimes 2} + \frac{m_\sys^2}{M^2} \frac{\avg{\vect{k}_\bath^2}_\bath}{d} \matr{1}  \:.
\end{equation}
The average of the scalar product $\vect{k}^2$ is just the trace of the result~\eqref{eq:momentum-dyadic-bath-avg}
\begin{equation}\label{eq:momentum-scalar-bath-avg}
\avg{\vect{k}^2}_\bath = \frac{m_\bath^2}{M^2} \vect{k}_\sys^2 + \frac{m_\sys^2}{M^2} \avg{\vect{k}_\bath^2}_\bath  \:.
\end{equation}
Finally, substituting Eqs.~\eqref{eq:momentum-dyadic-bath-avg} and~\eqref{eq:momentum-scalar-bath-avg} into Eq.~\eqref{eq:second-rate-moment-6} and using Eq.~\eqref{eq:cross-section-quad-to-transfer} leads to the result
\begin{equation}\label{eq:second-rate-moment-final}
\matr{A}^{(2)} = \alpha_{\rm q\para} \frac{m_\bath^2}{M^2} \vect{k}_\sys^{\otimes 2}  
 + \alpha_{\rm q\perp} \frac{m_\bath^2}{M^2} \frac{\vect{k}_\sys^2\matr{1} - \vect{k}_\sys^{\otimes 2}}{d-1} 
 + 2\alpha_{\rm tr} \frac{m_\sys^2}{M^2} \kth^2 \matr{1}  \:,
\end{equation}
where $\kth$ is a compact notation for the characteristic wave number of the bath:
\begin{equation}\label{eq:def-bath-wavenumber}
\kth^2 = \frac{\avg{\vect{k}_\bath^2}_\bath}{d}  \:.
\end{equation}
In the special case of a classical gas described at equilibrium by the Maxwell-Boltzmann velocity distribution~\cite{Huang1987, Hornberger2003b, Hornberger2009, Vacchini2009}, the characteristic wave number is given by
\begin{equation}\label{eq:bath-wavenumber-thermal}
\kth^2 = \frac{m_\bath}{\hbar^2} \kbol T  \:.
\end{equation}

\subsection{Caldeira-Leggett form of the master equation}\label{sec:caldeira-leggett}
One returns to the expansion~\eqref{eq:redfieldian-expansion-2} of the dissipator in order to substitute the expressions found in Eqs.~\eqref{eq:first-rate-moment-final} and~\eqref{eq:second-rate-moment-final}.
It is useful to assume that the collisional rates $\alpha_\mu(k)$ are smooth enough functions of $k$ to get them out of the commutators $[\op{\vect{r}}, \cdot]$.
The result is
\begin{equation}\label{eq:caldeira-leggett-1}\begin{split}
\pder{\op{\rho}_\sys}{t} - \supop{L}_\sys\op{\rho}_\sys & = -\I\alpha_{\rm tr}\frac{m_\bath}{M} [\op{r}_i, \op{k}_{\sys,i}\op{\rho}_\sys]  \\
 & - \frac{\alpha_{\rm q\para}}{2} \frac{m_\bath^2}{M^2} [\op{r}_i, [\op{r}_j, \op{k}_{\sys,i}\op{k}_{\sys,j} \op{\rho}_\sys]]  \\
 & - \frac{\alpha_{\rm q\perp}}{2} \frac{m_\bath^2}{M^2} [\op{r}_i, [\op{r}_j, \frac{\op{\vect{k}}_\sys^2\delta_{ij} - \op{k}_{\sys,i}\op{k}_{\sys,j}}{d-1} \op{\rho}_\sys]]  \\
 & - \alpha_{\rm tr} \frac{m_\sys^2}{M^2} \kth^2 [\op{r}_i, [\op{r}_i, \op{\rho}_\sys]]  \:.
\end{split}\end{equation}
This equation has a form similar to the Caldeira-Leggett master equation~\cite{Caldeira1983b, Breuer2002, Diosi1993b, Vacchini2009, Hornberger2009, Kamleitner2010}.
As a reminder, the Caldeira-Leggett equation was derived for a different model in which the particle is coupled to a thermal bath of harmonic oscillators.
In fact, Eq.~\eqref{eq:caldeira-leggett-1} reduces to the usual Caldeira-Leggett equation in the limit $m_\sys\gg m_\bath$ for the Brownian motion, because the terms with the coefficients $\alpha_{\rm q\para}$ and $\alpha_{\rm q\perp}$ are then negligible.
\par The first term on the right-hand side of Eq.~\eqref{eq:caldeira-leggett-1} can be interpreted as friction, the second term as decoherence parallel to the direction of motion, and the third term as decoherence in the transverse direction.
The last term is the additional contribution to decoherence due to the thermal motion of the scatterers.
This contribution is isotropic since it acts equally in all directions.
\par One can check that the evolution equations of the moments $\tavg{E_{\vect{k}_\sys}}$ and $\tavg{\vect{k}_\sys}$ predicted by Eq.~\eqref{eq:caldeira-leggett-1} indeed reduce to Eqs.~\eqref{eq:energy-evolution-final} and~\eqref{eq:momentum-evolution-final} obtained in Sec.~\ref{sec:energy-transfer}.
In particular, all the terms of Eq.~\eqref{eq:caldeira-leggett-1} contribute to the evolution of $\tavg{E_{\vect{k}_\sys}}$, but only the first one, namely the friction term, contributes to the evolution of $\tavg{\vect{k}_\sys}$.

\subsubsection{Strongly forward scattering}\label{sec:forward-scattering}
Here, one simplifies Eq.~\eqref{eq:caldeira-leggett-1} by accounting for the strong directionality of the cross section in the forward direction.
This directionality is especially expected for high-energy particles.
The assumption of strong forward scattering typically implies that the transfer cross section is much smaller than the total cross section, as in Eq.~\eqref{eq:kramers-moyal-condition-2}.
Under this assumption, the moments of the cross section defined in Eqs.~\eqref{eq:cross-section-quad-moment-para} and~\eqref{eq:cross-section-quad-moment-perp} can be approximated as
\begin{equation}\label{eq:cs-quad-approx-1}\begin{cases}
\sigma_{\rm q\para}(k) = \sigma(k) \avg{(1-\cos\theta)^2} \simeq \tfrac{1}{4} \sigma(k) \avg{\theta^4} \:,\\
\sigma_{\rm q\perp}(k) = \sigma(k) \avg{\sin^2\theta}     \simeq \sigma(k) \avg{\theta^2}   \:,
\end{cases}\end{equation}
where $\tavg{\cdot}$ denotes the average weighted by the differential cross section $\textstyle\der{\sigma}{\Omega}$.
Indeed, the scattering angle $\theta$ is relatively small compared to $1$.
The fourth angular moment can be related to the kurtosis $\kappa$ as $\tavg{\theta^4}=\kappa\tavg{\theta^2}^2$.
If the differential cross section is mesokurtic, i.e., similar to a normal distribution, then one has $\kappa=3$.
Moreover, since it is known that the transfer cross section is approximately $\sigma_{\rm tr}(k)\simeq\tfrac{1}{2}\sigma(k)\tavg{\theta^2}$, one can write
\begin{equation}\label{eq:cs-quad-approx-2}\begin{cases}
\sigma_{\rm q\para}(k) \simeq \kappa\frac{\sigma_{\rm tr}(k)^2}{\sigma(k)} \ll \kappa\sigma_{\rm tr}(k)  \:,\\
\sigma_{\rm q\perp}(k) \simeq 2\sigma_{\rm tr}(k)  \:.
\end{cases}\end{equation}
If $\kappa$ is of the order of $3$, then Eq.~\eqref{eq:cs-quad-approx-2} shows that the longitudinal moment is typically much smaller than the transverse moment ($\sigma_{\rm q\para}(k)\ll\sigma_{\rm q\perp}(k)$), and can be neglected.
Therefore, reasonable values for the cross-sectional moments~\eqref{eq:cs-quad-approx-2} could be
\begin{equation}\label{eq:cs-quad-approx-final}\begin{cases}
\sigma_{\rm q\para}(k) \simeq 0  \:,\\
\sigma_{\rm q\perp}(k) \simeq 2\sigma_{\rm tr}(k)  \:,
\end{cases}\end{equation}
In this way, the moments still satisfy the constraint given by Eq.~\eqref{eq:cross-section-quad-to-transfer}.
Note that assuming $\sigma_{\rm q\para}(k)\simeq 0$ completely discards the longitudinal momentum diffusion.
This can be geometrically understood from Fig.~\ref{fig:binary-collision-moments}.
Indeed, when $\abs{\tilde{u}(\vect{q})}^2$ is concentrated at $\vect{q}=\vect{0}$, the outgoing momenta are constrained to a very thin region which mostly extends in the transverse directions.
Accordingly, the momentum diffuses much more slowly in the longitudinal direction.
\par Letting $\alpha_{\rm q\para}$ tend to zero in Eq.~\eqref{eq:caldeira-leggett-1} leads to
\begin{equation}\label{eq:caldeira-leggett-2}\begin{split}
\pder{\op{\rho}_\sys}{t} - \supop{L}_\sys\op{\rho}_\sys & = -\I\alpha_{\rm tr}\frac{m_\bath}{M} [\op{r}_i, \op{k}_{\sys,i}\op{\rho}_\sys]  \\
 & - \alpha_{\rm tr} \frac{m_\bath^2}{M^2} [\op{r}_i, [\op{r}_j, \frac{\op{\vect{k}}_\sys^2\delta_{ij} - \op{k}_{\sys,i}\op{k}_{\sys,j}}{d-1} \op{\rho}_\sys]]  \\
 & - \alpha_{\rm tr} \frac{m_\sys^2}{M^2} \kth^2 [\op{r}_i, [\op{r}_i, \op{\rho}_\sys]]  \:.
\end{split}\end{equation}
Let us focus on the second term of Eq.~\eqref{eq:caldeira-leggett-2} associated to the deflection of the particle in the directions perpendicular to the direction of motion.
To better understand this term, it is useful to express it in terms of the angular momentum operator, which is defined in arbitrary dimension as
\begin{equation}\label{eq:def-angular-momentum}
\op{L}_{ij} = \op{r}_i\op{k}_{\sys,j} - \op{r}_j\op{k}_{\sys,i}  \:.
\end{equation}
The operator $\op{L}_{ij}$ is the generator of the rotation in the plane $ij$, and thus acts in both position and momentum spaces.
\par The second term of Eq.~\eqref{eq:caldeira-leggett-2} can be related to $\op{L}_{ij}$ by the nontrivial property
\begin{equation}\label{eq:angular-momentum-property}\begin{split}
[\op{r}_i, [\op{r}_j, (\op{\vect{k}}_\sys^2\delta_{ij} - \op{k}_{\sys,i}\op{k}_{\sys,j}) \op{\rho}_\sys]] & = 
 \frac{1}{2}[\op{L}_{ij}, [\op{L}_{ij}, \op{\rho}_\sys]]  \\
 & - \I(d-1) [\op{r}_i, \op{k}_{\sys,i} \op{\rho}_\sys]  \:,
\end{split}\end{equation}
which is proved in Appendix~\ref{app:decoherence-on-sphere} under the approximation~\eqref{eq:diagonality-approx}.
The interest of Eq.~\eqref{eq:angular-momentum-property} is the separation of the purely transverse contribution to decoherence (the first term on the right-hand side) from the residual contribution to the mean energy (the second term).
Substituting Eq.~\eqref{eq:angular-momentum-property} into Eq.~\eqref{eq:caldeira-leggett-2} yields
\begin{equation}\label{eq:caldeira-leggett-3}\begin{split}
\pder{\op{\rho}_\sys}{t} - \supop{L}_\sys\op{\rho}_\sys & = -\I\alpha_{\rm tr}\frac{m_\sys m_\bath}{M^2} [\op{r}_i, \op{k}_{\sys,i}\op{\rho}_\sys]  \\
 & - \frac{\alpha_{\rm tr}}{2} \frac{m_\bath^2}{M^2} \frac{1}{d-1} [\op{L}_{ij}, [\op{L}_{ij}, \op{\rho}_\sys]]  \\
 & - \alpha_{\rm tr} \frac{m_\sys^2}{M^2} \kth^2 [\op{r}_i, [\op{r}_i, \op{\rho}_\sys]]  \:.
\end{split}\end{equation}
The first and third terms on the right-hand side of Eq.\ \eqref{eq:caldeira-leggett-3} are respectively the friction and decoherence terms, which are well known in the Caldeira-Leggett master equation~\cite{Caldeira1983b, Breuer2002, Diosi1993b, Vacchini2009, Hornberger2009, Kamleitner2010}.
However, the second term represents an additional contribution to the transverse decoherence, which is the central result of this paper.
This term has the effect of rotating the wave packet by an infinitesimal angle in a random direction under the impact of collisions with the scatterers.
\par Furthermore, one notices that the transport parameters in Eq.\ \eqref{eq:caldeira-leggett-3} merely depend on the transfer cross section\ \eqref{eq:def-transfer-cross-section}.
The transfer cross section is thus the main relevant phenomenological parameter governing the transport of fast particles in matter~\cite{Landau1967}.

\subsubsection{Wigner representation}\label{sec:wigner-representation}
In this section, the Wigner transform of Eq.~\eqref{eq:caldeira-leggett-3} is derived.
As a reminder, the Wigner transform can be defined in either the position or the momentum basis as follows~\cite{Wigner1932, Moyal1949a, Basdevant2002, Cohen2020-vol3}
\begin{equation}\label{eq:def-wigner-general}\begin{split}
\wigner{\op{A}} & = \int_{\mathbb{R}^d} \frac{\D\vect{s}}{(2\pi)^d} \bra{\vect{k}_\sys+\tfrac{\vect{s}}{2}} \op{A} \ket{\vect{k}_\sys-\tfrac{\vect{s}}{2}} \E^{\I\vect{s}\cdot\vect{r}}  \\
 & = \int_{\mathbb{R}^d} \frac{\D\vect{s}}{(2\pi)^d} \bra{\vect{r}+\tfrac{\vect{s}}{2}} \op{A} \ket{\vect{r}-\tfrac{\vect{s}}{2}} \E^{-\I\vect{k}_\sys\cdot\vect{s}}  \:.
\end{split}\end{equation}
In particular, the Wigner transform of the density matrix gives the Wigner function
\begin{equation}\label{eq:def-wigner-function}
f_\sys(\vect{r},\vect{k}_\sys) = \wigner{\op{\rho}_\sys}  \:.
\end{equation}
The function $f_\sys(\vect{r},\vect{k}_\sys)$ is a real function of the position $\vect{r}$ and the momentum $\vect{k}_\sys$.
This function is also referred to as quasi-probability distribution because of its similarity to the classical phase-space distribution.
In particular, it is normalized according to~\footnote{The position integral is restricted to the region $\mathcal{V}$, because of the periodic boundary conditions on the particle wave function. See also Sec.~\ref{sec:general-properties}.}
\begin{equation}\label{eq:wigner-function-normalization}
\int_{\mathcal{V}} \D\vect{r} \int_{\mathbb{R}^d} \D\vect{k}_\sys f_\sys(\vect{r},\vect{k}_\sys) = 1  \:.
\end{equation}
However, in contrast to a usual probability distribution, $f_\sys(\vect{r},\vect{k}_\sys)$ may be negative, typically in the presence of quantum interferences.
Before applying the Wigner transform to Eq.~\eqref{eq:caldeira-leggett-3}, it is useful to consider the following transform of the commutators:
\begin{equation}\label{eq:wigner-commutators}\begin{cases}
\wigner{[\op{\vect{r}}, \op{\rho}_\sys]}      =  \I\grad_{\vect{k}_\sys} f_\sys(\vect{r},\vect{k}_\sys)  \:,\\
\wigner{[\op{\vect{k}}_\sys, \op{\rho}_\sys]} = -\I\grad_{\vect{r}} f_\sys(\vect{r},\vect{k}_\sys)       \:.
\end{cases}\end{equation}
More generally, each commutator, $[\op{\vect{r}}, \cdot]$ or $[\op{\vect{k}}_\sys, \cdot]$, corresponds to a multiplication by a gradient, respectively $\I\grad_{\vect{k}_\sys}$ or $-\I\grad_{\vect{r}}$, in the Wigner representation.
In addition, the Wigner transform of the double commutator with $\op{L}_{ij}$ is given by
\begin{equation}\label{eq:wigner-spherical-decoherence}
\wigner{\frac{1}{2}[\op{L}_{ij}, [\op{L}_{ij}, \op{\rho}_\sys]]} = -\lapl_{\perp\vect{k}_\sys} f_\sys(\vect{r},\vect{k}_\sys)  \:,
\end{equation}
where $\lapl_{\perp\vect{k}_\sys}$ is the Laplace-Beltrami operator on the spherical submanifold of $\mathbb{R}^d$.
The result~\eqref{eq:wigner-spherical-decoherence} is proved at the end of Appendix~\ref{app:decoherence-on-sphere}.
Using Eqs.~\eqref{eq:wigner-commutators} and~\eqref{eq:wigner-spherical-decoherence}, the Wigner representation of Eq.~\eqref{eq:caldeira-leggett-3} reads
\begin{equation}\label{eq:fokker-planck-fast}
\pder{f_\sys}{t} + \vect{v}_\sys\cdot\grad_{\vect{r}}f_\sys = \eta \grad_{\vect{k}_\sys}\cdot\big(\vect{k}_\sys f_\sys\big) 
 + \gamma \lapl_{\perp\vect{k}_\sys} f_\sys 
 + \xi \lapl_{\vect{k}_\sys} f_\sys  \:,
\end{equation}
where $\vect{v}_\sys=\tfrac{\hbar\vect{k}_\sys}{m_\sys}$ is the particle velocity, $\lapl_{\vect{k}_\sys}$ denotes the standard Laplace operator, and the transport coefficients are defined as
\begin{equation}\label{eq:kramers-parameters}
\eta = \frac{m_\sys m_\bath}{M^2} \alpha_{\rm tr}   \:,\quad
\gamma = \frac{m_\bath^2}{M^2} \frac{\alpha_{\rm tr}}{d-1}  \:,\quad
\xi = \frac{m_\sys^2}{M^2} \alpha_{\rm tr} \kth^2  \:.
\end{equation}
The parameter $\gamma$ will be referred to as the directional diffusivity, or the transverse diffusivity, and $\xi$ as the momentum diffusivity induced by the bath.
It is remarkable that all the parameters in Eq.~\eqref{eq:kramers-parameters} are determined by the single collisional parameter $\alpha_{\rm tr}$.
Between $\eta$ and $\xi$, this relationship can be interpreted as a consequence of the fluctuation-dissipation theorem~\cite{Huang1987, Breuer2002}
\begin{equation}\label{eq:fluctuation-dissipation}
\frac{\xi}{\eta} = \frac{m_\sys\kbol T}{\hbar^2}  \:,
\end{equation}
assuming $\kth$ is given by Eq.~\eqref{eq:bath-wavenumber-thermal}.
\par Equations of the form of Eq.~\eqref{eq:fokker-planck-fast} are generally called Fokker-Planck equations~\cite{Caldeira1983b, Diosi1995, Vacchini2009, Breuer2002} or, occasionally, Kramers equations~\cite{Kamleitner2010}.
Equations similar to Eq.~\eqref{eq:fokker-planck-fast} were found in Refs.~\cite{Wax1998, Cheng1999} for light beams in the paraxial approximation, but without the friction term.
\par Finally, it is worth noting that in the context of fast particles which are much faster than the scatterers ($\tavg{\vect{v}_\sys^2}\gg\tavg{\vect{v}_\bath^2}$), the contribution to the angular diffusion from the term $\xi\lapl_{\vect{k}_\sys}f_\sys$ can be neglected at the beginning of the propagation.

\section{Coherence length}\label{sec:coherence-length}
When studying decoherence on a given density matrix, it can be useful to determine the characteristic length beyond which quantum coherence disappears.
This coherence manifests itself off the diagonal of the density matrix $\rho_\sys(\vect{r},\dual{\vect{r}})$, that is for $\vect{r}\neq\dual{\vect{r}}$.
In practice, the spatial extent of this coherent region can be estimated by the variance of $\abs{\rho_\sys(\vect{r},\dual{\vect{r}})}^2$ in the separation variable $\vect{s}=\vect{r}-\dual{\vect{r}}$, as proposed in Ref.~\cite{Barnett2000}.
Furthermore, it is also possible to characterize the ellipsoidal shape of the coherent region using the covariance matrix.
This idea leads to the definition of the coherence length matrix:
\begin{equation}\label{eq:def-coherence-length-matrix}
\matr{\lcoh}^2 = \frac{\iint_{\mathcal{V}^2} (\vect{r}-\dual{\vect{r}})^{\otimes 2} \abs{\rho_\sys(\vect{r}, \dual{\vect{r}})}^2 \D\vect{r} \D\dual{\vect{r}}}{2\iint_{\mathcal{V}^2} \abs{\rho_\sys(\vect{r}, \dual{\vect{r}})}^2 \D\vect{r} \D\dual{\vect{r}}}  \:.
\end{equation}
The factor of two in the denominator of Eq.~\eqref{eq:def-coherence-length-matrix} is needed to make this definition consistent with the variance of a pure state $\rho_\sys(\vect{r},\dual{\vect{r}})=\psi(\vect{r})\cc{\psi}(\dual{\vect{r}})$, as pointed out in Ref.~\cite{Barnett2000}.
Definition~\eqref{eq:def-coherence-length-matrix} can also be written using the quantum operator notations as
\begin{equation}\label{eq:coherence-length-matrix-formal-1}
[\matr{\lcoh}^2]_{ij} = \frac{\Tr_\sys(\op{\rho}_\sys^2\op{r}_i\op{r}_j - \op{\rho}_\sys\op{r}_i\op{\rho}_\sys\op{r}_j)}{\Tr_\sys(\op{\rho}_\sys^2)}  \:,
\end{equation}
but also as
\begin{equation}\label{eq:coherence-length-matrix-formal-2}
[\matr{\lcoh}^2]_{ij} = -\frac{\Tr_\sys([\op{r}_i, \op{\rho}_\sys][\op{r}_j, \op{\rho}_\sys])}{2\Tr_\sys(\op{\rho}_\sys^2)}  \:,
\end{equation}
or even, using the property $\Tr([\op{A},\op{B}]\op{C})=-\Tr(\op{B}[\op{A},\op{C}])$, as
\begin{equation}\label{eq:coherence-length-matrix-formal-3}
[\matr{\lcoh}^2]_{ij} = \frac{\Tr_\sys(\op{\rho}_\sys [\op{r}_i, [\op{r}_j, \op{\rho}_\sys]])}{2\Tr_\sys(\op{\rho}_\sys^2)}  \:.
\end{equation}
In addition, the coherence length is related to the Wigner function of the particle by
\begin{equation}\label{eq:coherence-length-matrix-wigner-1}
\matr{\lcoh}^2 = \frac{\iint \grad_{\vect{k}_\sys}f_\sys(\vect{r},\vect{k}_\sys)\otimes\grad_{\vect{k}_\sys}f_\sys(\vect{r},\vect{k}_\sys) \D\vect{r}\D\vect{k}_\sys}{2\iint f_\sys(\vect{r},\vect{k}_\sys)^2 \D\vect{r}\D\vect{k}_\sys}  \:.
\end{equation}
Expression~\eqref{eq:coherence-length-matrix-wigner-1} comes directly from Eq.~\eqref{eq:coherence-length-matrix-formal-2} using Eq.~\eqref{eq:wigner-commutators} and the fact that
\begin{equation}
\Tr(\op{\rho}_\sys^2) = (2\pi)^d \iint f_\sys(\vect{r},\vect{k}_\sys)^2 \D\vect{r}\D\vect{k}_\sys  \:.
\end{equation}
There is yet another writing for the coherence length in terms of the Hessian matrix of the Wigner function, which reads
\begin{equation}\label{eq:coherence-length-matrix-wigner-2}
\matr{\lcoh}^2 = -\frac{\iint f_\sys(\vect{r},\vect{k}_\sys)\grad_{\vect{k}_\sys}\otimes\grad_{\vect{k}_\sys}f_\sys(\vect{r},\vect{k}_\sys) \D\vect{r}\D\vect{k}_\sys}{2\iint f_\sys(\vect{r},\vect{k}_\sys)^2 \D\vect{r}\D\vect{k}_\sys}  \:.
\end{equation}
This expression can be obtained from the integration by part of Eq.~\eqref{eq:coherence-length-matrix-wigner-1}, or by analogy from Eq.~\eqref{eq:coherence-length-matrix-formal-3}.
\par Expressions~\eqref{eq:coherence-length-matrix-wigner-1} and~\eqref{eq:coherence-length-matrix-wigner-2} show that the coherence length is larger when the Wigner function $f_\sys(\vect{r},\vect{k}_\sys)$ strongly varies in momentum space.
In particular, for a very peaked distribution such as for a plane wave, the coherence length can be infinite.
Note that the coherence length is also infinite for any discrete superposition of plane waves.
As the momentum diffusion takes place, due to the last two terms on the right-hand side of Eq.~\eqref{eq:fokker-planck-fast}, the momentum distribution smooths out on the continuum and its gradient diminishes.
Accordingly, the coherence length $\matr{\lcoh}^2$ is expected to decrease in time.
\par Furthermore, it should be noted that the definition~\eqref{eq:def-coherence-length-matrix} implicitly assumes that the integral converges.
This is not obvious, because the density matrix element given by Eq.~\eqref{eq:redfield-pos-basis-sol} in the general case does not vanish to zero for $\norm{\vect{r}-\dual{\vect{r}}}\rightarrow\infty$, but instead tends to a finite value.
This is due to the saturation of the decoherence rate observed in Eq.~\eqref{eq:decoherence-saturation}.
In that case, the coherence length defined as Eq.~\eqref{eq:def-coherence-length-matrix} would be infinite.
In fact, this issue is circumvented by the Kramers-Moyal expansion performed in Sec.~\ref{sec:kramers-moyal}, because then the decoherence rate $F(\vect{r}-\dual{\vect{r}})$ increases quadratically for $\norm{\vect{r}-\dual{\vect{r}}}\rightarrow\infty$ without any upper bound.
In this case, the density matrix element~\eqref{eq:redfield-pos-basis-sol} tends to zero for $\norm{\vect{r}-\dual{\vect{r}}}\rightarrow\infty$, and the coherence length is well defined.
Therefore, regarding the calculation of the coherence length from Eq.~\eqref{eq:def-coherence-length-matrix}, the Kramers-Moyal expansion amounts to neglecting the nonzero asymptotic value of $\rho_\sys(\vect{r},\dual{\vect{r}})$ for $\norm{\vect{r}-\dual{\vect{r}}}\rightarrow\infty$.

\subsection{Relation to the momentum variance}\label{sec:momentum-variance}
The distribution predicted by the Fokker-Planck equation~\eqref{eq:fokker-planck-fast} in the momentum space is typically of Gaussian nature, as it will be seen later in more details.
This is a fundamental feature of diffusion equations, and can be interpreted, in the context of random processes, as a consequence of the central limit theorem.
An important consequence of this Gaussian profile is that the coherence length turns out to be directly related to the variance of the momentum, as shown in this section.
First, one assumes that the Wigner function has the form of a multivariate Gaussian distribution
\begin{equation}\label{eq:general-gaussian}
f_\sys(\vect{r},\vect{k}_\sys) = \frac{1}{V (2\pi)^{\frac{d}{2}} \det\matr{K}} \E^{-\frac{1}{2}\tran{(\Delta\vect{k}_\sys)}\cdot\matr{K}^{-2}\cdot\Delta\vect{k}_\sys}  \:,
\end{equation}
where $\Delta\vect{k}_\sys=\vect{k}_\sys-\tavg{\vect{k}_\sys}$ is the centered momentum, and $\matr{K}^2$ is the momentum covariance matrix.
The distribution~\eqref{eq:general-gaussian} reduces to the equilibrium distribution for $\tavg{\vect{k}_\sys}=\vect{0}$ and $\matr{K}^2=\tfrac{m_\sys}{\hbar^2}\kbol T\matr{1}$.
The distribution~\eqref{eq:general-gaussian} is normalized according to Eq.~\eqref{eq:wigner-function-normalization}, and its covariance matrix is given by
\begin{equation}\label{eq:gaussian-coh-variance}
\int_{\mathcal{V}} \D\vect{r} \int_{\mathbb{R}^d} \D\vect{k}_\sys (\Delta\vect{k}_\sys)^{\otimes 2} f_\sys(\vect{r},\vect{k}_\sys) = \matr{K}^2  \:.
\end{equation}
Using the gradient of $f_\sys$ with respect to the momentum
\begin{equation}\label{eq:gaussian-coh-gradient}
\grad_{\vect{k}_\sys}f_\sys = -\matr{K}^{-2}\cdot\Delta\vect{k}_\sys f_\sys  \:,
\end{equation}
the integral in the numerator of Eq.~\eqref{eq:coherence-length-matrix-wigner-1} becomes
\begin{equation}\label{eq:gaussian-coh-step-1}
\int \grad_{\vect{k}_\sys}f_\sys\otimes\grad_{\vect{k}_\sys}f_\sys \D\vect{k}_\sys
 = \int \matr{K}^{-2}\cdot(\Delta\vect{k}_\sys)^{\otimes 2}\cdot\matr{K}^{-2} f_\sys^2 \D\vect{k}_\sys  \:.
\end{equation}
The right-hand side of Eq.~\eqref{eq:gaussian-coh-step-1} can be simplified with the property
\begin{equation}\label{eq:gaussian-coh-step-2}
\int (\Delta\vect{k}_\sys)^{\otimes 2} f_\sys^2 \D\vect{k}_\sys = \frac{\matr{K}^2}{2} \int f_\sys^2 \D\vect{k}_\sys  \:,
\end{equation}
which results from the fact that the variance of $f_\sys^2$ is half the variance of $f_\sys$ according to Eq.~\eqref{eq:general-gaussian}.
Therefore, substituting Eq.~\eqref{eq:gaussian-coh-step-2} into Eq.~\eqref{eq:gaussian-coh-step-1} and dividing both sides by $\textstyle 2\int f_\sys^2\D\vect{k}_\sys$ leads to
\begin{equation}\label{eq:gaussian-coh-step-3}
\matr{\lcoh}^2 = \frac{\int \grad_{\vect{k}_\sys}f_\sys\otimes\grad_{\vect{k}_\sys}f_\sys \D\vect{k}_\sys}{2\int f_\sys^2 \D\vect{k}_\sys}
 = \frac{\matr{K}^{-2}}{4}  \:,
\end{equation}
or, in simpler terms:
\begin{equation}\label{eq:variance-product-1}
\matr{\lcoh}^2\cdot\matr{K}^2 = \frac{\matr{1}}{4}  \:.
\end{equation}
The result~\eqref{eq:variance-product-1} can also be expressed with the rescaled momentum covariance matrix $\matr{P}^2=\hbar^2\matr{K}^2$ as
\begin{equation}\label{eq:variance-product-2}
\matr{\lcoh}^2\cdot\matr{P}^2 = \frac{\hbar^2}{4}\matr{1}  \:.
\end{equation}
Expressions~\eqref{eq:variance-product-1} and~\eqref{eq:variance-product-2} show that, for Gaussian states, the coherence length can be directly deduced from the knowledge of the momentum variance.
In addition, they confirm that the shorter the spatial coherence length, the larger the momentum variance.
This duality between position and momentum is strongly reminiscent of Heisenberg's uncertainty principle~\cite{Heisenberg1927, Messiah1961, Basdevant2002, Hall2013} for the given direction $i$
\begin{equation}\label{eq:uncertainty-relation}
\avg{\Delta\op{r}_i^2} \avg{\Delta\op{p}_i^2} \geq \frac{\hbar^2}{4}  \:,
\end{equation}
where $\Delta\op{r}_i=\op{r}_i-\tavg{\op{r}_i}$ and $\Delta\op{p}_i=\op{p}_i-\tavg{\op{p}_i}$ are the centered position and momentum, respectively.
\par The existence of an uncertainty relation for the coherence length follows from the fact that, according to the definition~\eqref{eq:def-wigner-general}, the coherence function $\rho_\sys(\vect{r}+\tfrac{\vect{s}}{2},\vect{r}-\tfrac{\vect{s}}{2})$ of variable $\vect{s}$ is related to the Wigner function $f_\sys(\vect{r},\vect{k}_\sys)$ by the Fourier transform with respect to $\vect{s}$, in the same way the position-space and the momentum-space wave functions are related in quantum mechanics.
Therefore, Eq.~\eqref{eq:variance-product-2} could be generalized to an inequality for non-Gaussian states.
However, such a generalization is not needed in this paper because the state will be assumed to be Gaussian.
In this context, the strict equality~\eqref{eq:variance-product-2} at the lower bound of Eq.~\eqref{eq:uncertainty-relation} implies that the decoherence process in particle detectors could, in principle, resolve the position of the particle at the quantum limit for an ideal measurement~\cite{Braginsky1992}.

\subsection{Evolution of the momentum distribution}\label{sec:momentum-evolution}
In view of obtaining approximate expressions for the coherence lengths in the longitudinal and transverse directions, the property~\eqref{eq:variance-product-1} will be exploited on estimates for the momentum variances, starting from the initial condition~\eqref{eq:initial-pure-state}, and assuming the transport parameters~\eqref{eq:kramers-parameters} do not depend on the particle energy.
In this way, finding the explicit solution of the Fokker-Planck equation~\eqref{eq:fokker-planck-fast} is not necessary.
The interest of this approach is that the momentum variance, defined as
\begin{equation}\label{eq:def-total-momentum-variance}
K^2 = \Tr(\matr{K}^2) = \Var(\vect{k}_\sys) = \tavg{\vect{k}_\sys^2} - \tavg{\vect{k}_\sys}^2  \:,
\end{equation}
can be found from the moment equations~\eqref{eq:energy-evolution-final} and~\eqref{eq:momentum-evolution-final} derived in the general case, hence bypassing Eq.~\eqref{eq:fokker-planck-fast}.
This approach should nevertheless lead to results consistent with Eq.~\eqref{eq:fokker-planck-fast}.
The equation for the average momentum is the same as Eq.~\eqref{eq:momentum-evolution-final}, repeated here for convenience:
\begin{equation}\label{eq:momentum-mean-evolution}
\der{\avg{\vect{k}_\sys}}{t} = -\zeta \avg{\vect{k}_\sys}  \:.
\end{equation}
The solution of Eq.~\eqref{eq:momentum-mean-evolution}, subjected to the initial condition $\avg{\vect{k}_\sys(0)}=\vect{k}_{\sys,0}$, reads
\begin{equation}\label{eq:momentum-mean-sol}
\avg{\vect{k}_\sys} = \vect{k}_{\sys,0} \E^{-\zeta t}  \:.
\end{equation}
Since the wave number is related to the velocity by $\vect{v}_\sys=\tfrac{\hbar\vect{k}_\sys}{m_\sys}$, Eq.~\eqref{eq:momentum-mean-sol} leads after time integration to the average distance traveled by the particle in its initial direction
\begin{equation}\label{eq:traveled-distance}
L_\sys = \frac{v_{\sys,0}}{\zeta} (1 - \E^{-\zeta t})  \:,
\end{equation}
where $v_{\sys,0}$ is the initial velocity of the particle.
The total distance traveled by the particle, or range~\cite{Segre1953, Segre1977, Ziegler1985, Sigmund2006, Sigmund2014}, is thus $L_{\sys,\infty}=v_{\sys,0}/\zeta$.
For instance, the well-known range of an alpha particle of initial kinetic energy of $5\,\mathrm{MeV}$ in dry air under normal conditions ($20^\circ\mathrm{C}$, $1\,\mathrm{atm}$) is $L_{\sys,\infty}=3.5\,\mathrm{cm}$ \cite{Bragg1905, Rutherford1924, Livingston1937, Segre1977, Sigmund2006, Sigmund2014, ESTAR}.
\par It should be noted that, according to the definition~\eqref{eq:def-momentum-friction-param}, $\zeta$ can be related to the parameters~\eqref{eq:kramers-parameters} as
\begin{equation}\label{eq:zeta-param-relation}
\zeta = \frac{m_\bath}{M}\alpha_{\rm tr} = \eta + (d-1)\gamma  \:.
\end{equation}
Equation~\eqref{eq:zeta-param-relation} shows that $\zeta$ is related to the directional diffusivity $\gamma$, which is not obvious for a friction parameter.
This is due to the fact that the directional diffusion, which is described by the $\gamma$ term in Eq.~\eqref{eq:fokker-planck-fast}, geometrically contributes to enhance the relaxation of $\tavg{\vect{k}_\sys}$ in time.
When the incident particle is heavy ($m_\sys\gg m_\bath$), the parameter $\gamma$ is much smaller than $\zeta$ or $\eta$.
In this particular case, the parameters $\zeta$ and $\eta$ are approximately equal to each other.
\par Next to the average momentum, the equation for $\tavg{\vect{k}_\sys^2}$ is given by Eq.~\eqref{eq:energy-evolution-final}, rewritten here as
\begin{equation}\label{eq:energy-mean-evolution}
\der{\avg{\vect{k}_\sys^2}}{t} = -2\eta \avg{\vect{k}_\sys^2} + 2d\xi  \:,
\end{equation}
using the parameters~\eqref{eq:kramers-parameters}.
The solution of Eq.~\eqref{eq:energy-mean-evolution} reads as in Eq.~\eqref{eq:energy-thermalization}
\begin{equation}\label{eq:energy-mean-sol}
\avg{\vect{k}_\sys^2} = \left(k_{\sys,0}^2 - \frac{d\xi}{\eta}\right)\E^{-2\eta t} + \frac{d\xi}{\eta}  \:.
\end{equation}
Note that, in Eq.~\eqref{eq:energy-mean-sol}, the initial condition $\Var(\vect{k}_\sys)=0$ and thus $\tavg{\vect{k}_\sys^2}=k_{\sys,0}^2$ was used.
This results from the assumption that the initial state is a plane wave.
Combining Eqs.~\eqref{eq:momentum-mean-sol} and~\eqref{eq:energy-mean-sol} into Eq.~\eqref{eq:def-total-momentum-variance} leads to
\begin{equation}\label{eq:momentum-variance-final}
K^2 = k_{\sys,0}^2(\E^{-2\eta t} - \E^{-2\zeta t}) + \frac{d\xi}{\eta}(1 - \E^{-2\eta t})   \:.
\end{equation}
This result should be understood as the total variance of the momentum, which is the sum of the longitudinal and the transverse variances.
Since there is one longitudinal direction and $(d-1)$ transverse directions in $\mathbb{R}^d$, the total variance~\eqref{eq:momentum-variance-final} can be decomposed as follows
\begin{equation}\label{eq:momentum-variance-decomposition}
K^2 = K_\para^2 + (d-1)K_\perp^2  \:.
\end{equation}
In order to identify $K_\para$ and $K_\perp$ in Eq.~\eqref{eq:momentum-variance-final}, one way is to temporarily remove the contribution from the transverse diffusion by letting $\gamma=0$.
Doing so, the parameters $\zeta$ and $\eta$ become equal to each other according to Eq.~\eqref{eq:zeta-param-relation}, and the first term on the right-hand side of Eq.~\eqref{eq:momentum-variance-final} vanishes.
What remains should correspond to the isotropic contribution from the thermal bath, and what disappears should correspond to the contribution to the transverse variance.
Therefore, the sought decomposition reads
\begin{equation}\label{eq:def-momentum-variances}\begin{cases}
K_\para^2 = \frac{\xi}{\eta}(1 - \E^{-2\eta t})  \:,\\
K_\perp^2 = k_{\sys,0}^2\frac{\E^{-2\eta t} - \E^{-2\zeta t}}{d-1} + \frac{\xi}{\eta}(1 - \E^{-2\eta t})  \:.
\end{cases}\end{equation}
This result can be confirmed by more detailed calculations based on Eq.~\eqref{eq:fokker-planck-fast} which are not presented here.
\par
\begin{figure}[ht]%
\includegraphics{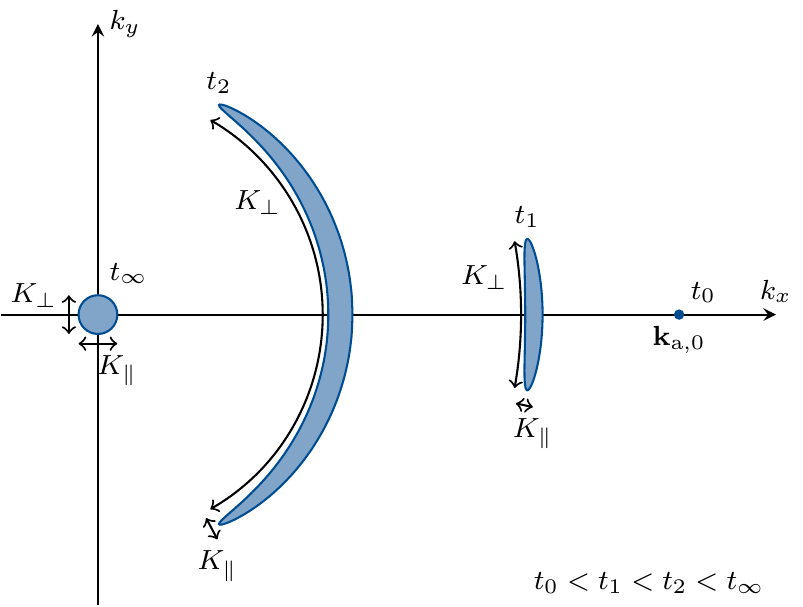}%
\caption{Schematic representation of the momentum distribution predicted by Eq.~\eqref{eq:fokker-planck-fast} at different times.
The geometric interpretation of the standard deviations $K_\para$ and $K_\perp$ given by Eq.~\eqref{eq:def-momentum-variances} is also highlighted.
The time $t_0=0$ corresponds to the initial condition, and $t_\infty$ to the equilibrium distribution.}%
\label{fig:momentum-distribution}%
\end{figure}

 The behavior of the momentum distribution satisfying Eq.~\eqref{eq:fokker-planck-fast} is schematically depicted over time in Fig.~\ref{fig:momentum-distribution} for the initial momentum $\vect{k}_{\sys,0}$ at time $t_0=0$.
This figure helps to geometrically interpret the parameters $K_\para$ and $K_\perp$ given by Eq.~\eqref{eq:def-momentum-variances}.
At time $t_1$, the distribution spreads more in the transverse direction than in the longitudinal direction under the effect of the $\gamma$ term in Eq.~\eqref{eq:fokker-planck-fast}.
The average momentum, $\tavg{\vect{k}_\sys}$, also gets smaller due to friction.
At the later time $t_2$, the angular aperture of the distribution increases until it forms a spherical cap around the point $\vect{k}_\sys=\vect{0}$.
At time $t_\infty$ after a while ($t_\infty\rightarrow\infty$), the distribution tends to the equilibrium Boltzmann distribution which is isotropic and centered at $\vect{k}_\sys=\vect{0}$.
\par The behavior depicted in Fig.~\ref{fig:momentum-distribution}, and especially the fact that $K_\perp$ increases faster than $K_\para$, can be checked in the short-time limit:
\begin{equation}\label{eq:momentum-variances-small-time}\begin{cases}
K_\para^2 \xrightarrow{t\rightarrow 0} 2\xi t  \:,\\
K_\perp^2 \xrightarrow{t\rightarrow 0} 2k_{\sys,0}^2\gamma t + 2\xi t  \:.
\end{cases}\end{equation}
In the second line of Eq.~\eqref{eq:momentum-variances-small-time}, one notices that the contribution of the term $k_{\sys,0}^2\gamma$ is much larger than that of $\xi$ for a fast particle ($v_{\sys,0}^2\gg\tavg{\vect{v}_\bath^2}$).
Indeed, the ratio of these parameters reads
\begin{equation}\label{eq:ratio-gamma-to-xi}
\frac{k_{\sys,0}^2\gamma}{\xi} = \frac{d}{d-1} \frac{v_{\sys,0}^2}{\avg{\vect{v}_\bath^2}}  \:,
\end{equation}
according to Eqs.~\eqref{eq:def-bath-wavenumber} and~\eqref{eq:kramers-parameters}.
Furthermore, the two variances in Eq.~\eqref{eq:def-momentum-variances} effectively converges to the same value in the long time limit:
\begin{equation}\label{eq:momentum-variances-large-time}
K_\para^2 \xrightarrow{t\rightarrow\infty} K_\perp^2 \xrightarrow{t\rightarrow\infty} \frac{\xi}{\eta} = \frac{m_\sys}{\hbar^2}\kbol T  \:.
\end{equation}
This is consistent with the regime of thermal equilibrium with the gas shown at time $t_\infty$ in Fig.~\ref{fig:momentum-distribution}.
Indeed, in this regime, the momentum distribution of the particle tends to the Boltzmann distribution~\eqref{eq:equilibrium-distrib} which exhibits the same variance in every individual direction, and whose value coincides with Eq.~\eqref{eq:momentum-variances-large-time}.

\subsection{Evolution of the coherence lengths}\label{sec:coherence-evolution}
\begin{figure}[ht]%
\includegraphics{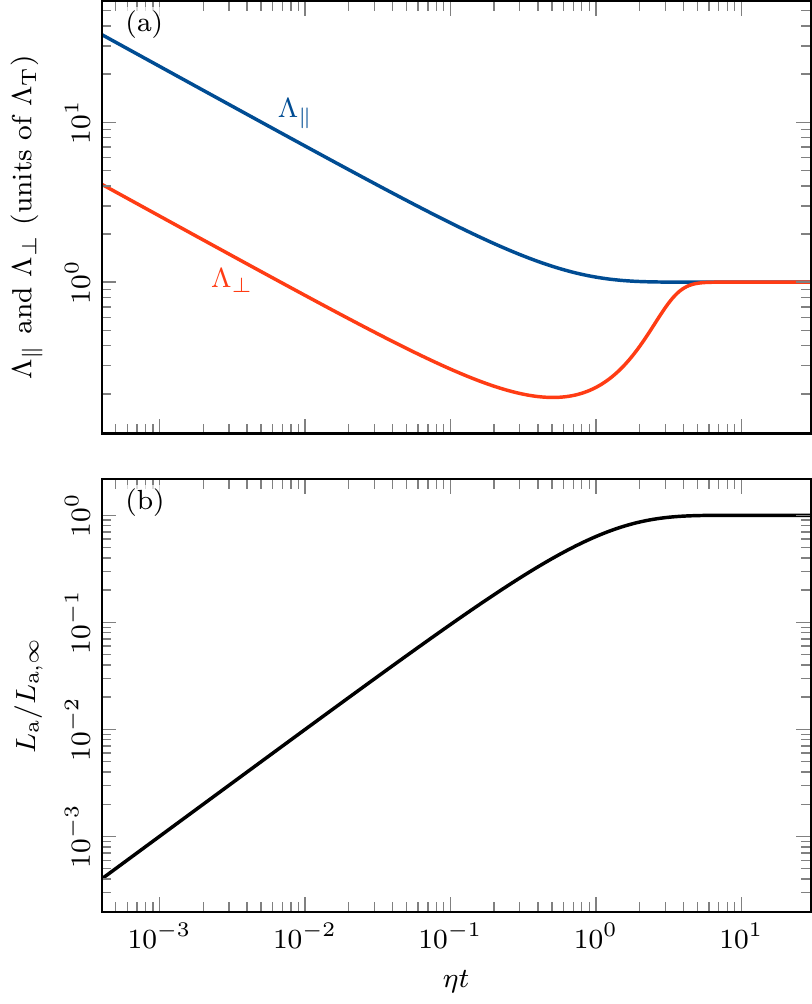}%
\caption{(a) Coherence lengths $\lcoh_\para$ and $\lcoh_\perp$ of the incident particle given by Eqs.~\eqref{eq:def-momentum-variances} and~\eqref{eq:def-para-perp-coherence-lengths} normalized by the equilibrium value $\lcoh_{\rm T}$ in Eq.~\eqref{eq:thermal-coherence-length}.
(b) Traveled distance in Eq.~\eqref{eq:traveled-distance} as a function of time.
The velocity ratio is set to $v_{\sys,0}/\tavg{\vect{v}_\bath^2}^{1/2}=7$ in accordance with Eqs.~\eqref{eq:alpha-initial-velocity} and~\eqref{eq:typical-electron-velocity}, and the mass ratio is $m_\sys/m_\bath=10^3$.
The curves of both panels depend little on the mass ratio for $m_\sys\gg m_\bath$.}%
\label{fig:coherence-lengths-alpha}%
\end{figure}

According to the property~\eqref{eq:variance-product-1}, the results~\eqref{eq:def-momentum-variances} leads to interesting approximations of the coherence lengths in the longitudinal and transverse direction:
\begin{equation}\label{eq:def-para-perp-coherence-lengths}
\lcoh_\para = \frac{1}{2K_\para}  \quad\text{and}\quad  \lcoh_\perp = \frac{1}{2K_\perp}  \:.
\end{equation}
The time evolution of the coherence lengths~\eqref{eq:def-para-perp-coherence-lengths} following Eq.~\eqref{eq:def-momentum-variances} is shown in Fig.~\ref{fig:coherence-lengths-alpha}(a) along with the traveled distance~\eqref{eq:traveled-distance} in Fig.~\ref{fig:coherence-lengths-alpha}(b).
Since $K_\para<K_\perp$ in Eq.~\eqref{eq:def-momentum-variances}, the coherence length in Eq.~\eqref{eq:def-para-perp-coherence-lengths} is typically smaller in the transverse direction than in the longitudinal direction ($\lcoh_\perp<\lcoh_\para$).
Therefore, the coherent wave packet can be thought of as an ellipsoid elongated in the direction of motion.
According to Eq.~\eqref{eq:momentum-variances-small-time}, the coherence lengths~\eqref{eq:def-para-perp-coherence-lengths} both behave in the short-time limit as the power law $t^{-1/2}$:
\begin{equation}\label{eq:coherence-lengths-small-time}\begin{cases}
\lcoh_\para \xrightarrow{t\rightarrow 0} \frac{1}{2} (2\xi t)^{-\frac{1}{2}}  \:,\\[8pt]
\lcoh_\perp \xrightarrow{t\rightarrow 0} \frac{1}{2} (2k_{\sys,0}^2\gamma t + 2\xi t)^{-\frac{1}{2}}  \:.
\end{cases}\end{equation}
This explains why, in log-log scale, the curves of $\lcoh_\para$ and $\lcoh_\perp$ look parallel in Fig.~\ref{fig:coherence-lengths-alpha}(a) for $\eta t\ll 1$.
An important consequence is that, at short times, the ratio of the coherence lengths keeps the constant value
\begin{equation}\label{eq:coherence-lengths-ratio}
\frac{\lcoh_\para}{\lcoh_\perp} \xrightarrow{t\rightarrow 0} \sqrt{1 + \frac{d}{d-1} \frac{v_{\sys,0}^2}{\avg{\vect{v}_\bath^2}}}  \:,
\end{equation}
which only depends on the velocities of the incident particle and the scatterers, and not on any of the transport parameters that were introduced in Eq.~\eqref{eq:kramers-parameters}.
This gives to Eq.~\eqref{eq:coherence-lengths-ratio} a universal nature.
Note, however, that Eq.~\eqref{eq:coherence-lengths-ratio} neglects the contribution to longitudinal decoherence of the second term on the right-hand side of Eq.~\eqref{eq:caldeira-leggett-1}.
This additional contribution could make the ratio~\eqref{eq:coherence-lengths-ratio} smaller.
\par Using Eq.~\eqref{eq:coherence-lengths-ratio}, the coherence length ratio can be estimated for an alpha particle of initial kinetic energy of $E_{\sys,0}=5\,\mathrm{MeV}$ as considered by Darwin and Mott~\cite{Darwin1929, Mott1929}.
This particular energy is the most common energy for an alpha particle generated by natural radioactive emitters such as uranium-238 ($E_\alpha=4.270\,\mathrm{MeV}$)~\cite{Be2006-vol3}, radium-226 ($E_\alpha=4.871\,\mathrm{MeV}$), radon-222 ($E_\alpha=5.590\,\mathrm{MeV}$), or polonium-210 ($E_\alpha=5.407\,\mathrm{MeV}$)~\cite{Be2008-vol4}.
The velocity of such an alpha particle is
\begin{equation}\label{eq:alpha-initial-velocity}
\frac{v_{\sys,0}}{c} = \frac{\sqrt{E_{\sys,0}(E_{\sys,0} + 2m_\sys c^2)}}{E_{\sys,0} + m_\sys c^2} \simeq 0.052  \:,
\end{equation}
using the alpha particle mass $m_\sys c^2=3727\,\mathrm{MeV}$ \cite{Tiesinga2021}, and relativistic effects are thus negligible.
The alpha particle is assumed to mainly interact with the electrons of the medium.
Indeed, the fact that the slowdown of a fast particle is mainly due to collisions with electrons is well known and is true for any material including ordinary gases such as air~\cite{Segre1953, Segre1977, Ziegler1985, Sigmund2006, Sigmund2014}.
In addition, the mean electron velocity in matter is roughly given by the Bohr velocity~\cite{Sigmund2006, Sigmund2014}
\begin{equation}\label{eq:typical-electron-velocity}
\sqrt{\avg{\vect{v}_\bath^2}} \simeq \alpha c  \:,
\end{equation}
where $\alpha\simeq 1/137$ is the fine-structure constant.
Substituting the numerical values of Eqs.~\eqref{eq:alpha-initial-velocity} and~\eqref{eq:typical-electron-velocity} into Eq.~\eqref{eq:coherence-lengths-ratio} leads to
\begin{equation}\label{eq:typical-coherence-ratio}
\frac{\lcoh_\para}{\lcoh_\perp} \overset{t\rightarrow 0}{\simeq} 8.7  \:.
\end{equation}
The longitudinal elongation of the coherent wave packet is therefore significant at the start of the propagation in the particle detector.
\par In the long-time limit ($\eta t\gg 1$), the two coherence lengths $\lcoh_\para$ and $\lcoh_\perp$ tend to the same thermal wavelength
\begin{equation}\label{eq:thermal-coherence-length}
\lcoh_{\rm T} = \frac{1}{2}\sqrt{\frac{\eta}{\xi}} = \frac{\hbar}{2\sqrt{m_\sys\kbol T}}  \:.
\end{equation}
This shows that the coherent wave packet takes a spherical shape at equilibrium, as for the gas particles.
In the special case of a thermal alpha particle at $T=300\,\mathrm{K}$, Eq.~\eqref{eq:thermal-coherence-length} gives the value $\lcoh_{\rm T}\simeq 10\,\mathrm{pm}$.
\par Probably one the most interesting points in Fig.~\ref{fig:coherence-lengths-alpha}(a) is that the transverse coherence length, $\lcoh_\perp$, drops below the thermal length before rising to reach it.
This results from the important effect of momentum diffusion in the transverse direction.
Indeed, according to Eqs.~\eqref{eq:momentum-variances-small-time} and~\eqref{eq:ratio-gamma-to-xi}, the transverse diffusion rate is much larger than the thermal one ($\gamma k_\sys^2\gg\xi$) for a fast incident particle.
Therefore, shortly after entering the detector, the momentum distribution of the incident particle gets more extended in the transverse direction than the thermal distribution itself, as it can be seen at $t_1$ and $t_2$ in Fig.~\ref{fig:momentum-distribution}.
In contrast, the longitudinal coherence length, $\lcoh_\para$, does not drop below the thermal wavelength~\eqref{eq:thermal-coherence-length} according to Fig.~\ref{fig:coherence-lengths-alpha}(a).
However, it is quite possible that, if the longitudinal decoherence term, namely the second term on the right-hand side of Eq.~\eqref{eq:caldeira-leggett-1}, is retained from the beginning, the longitudinal coherence length will also manifest such an undershoot.
\par Finally, in order to better interpret the transverse coherence length $\lcoh_\perp$ and its time evolution for a fast particle, it is convenient to define the angular variance of the momentum distribution as
\begin{equation}\label{eq:def-angular-variance}
\avg{\theta^2} = \frac{K_\perp^2}{\avg{\vect{k}_\sys^2}}  \:,
\end{equation}
where $K_\perp^2$ is given by Eq.~\eqref{eq:def-momentum-variances} and $\tavg{\vect{k}_\sys^2}$ by Eq.~\eqref{eq:energy-mean-sol}.
In the case of a fast incident particle ($v_{\sys,0}^2\gg\tavg{\vect{v}_\bath^2}$), the thermal contribution can be neglected ($\xi\rightarrow 0$), and Eq.~\eqref{eq:def-angular-variance} can be approximated by
\begin{equation}\label{eq:angular-variance-approx}
\avg{\theta^2} \simeq 2\gamma t  \:,
\end{equation}
and the transverse coherence length~\eqref{eq:coherence-lengths-small-time} behaves at short time as
\begin{equation}\label{eq:perp-coh-len-from-ang-var}
\lcoh_\perp \xrightarrow{t\rightarrow 0} \frac{1}{2k_{\sys,0}\avg{\theta^2}^{\frac{1}{2}}} = \frac{\lambda_{\sys,0}}{4\pi\avg{\theta^2}^{\frac{1}{2}}}  \:,
\end{equation}
where $\lambda_{\sys,0}=2\pi/k_{\sys,0}$ is the de Broglie wavelength of the particle at the entrance in the medium.
Equation~\eqref{eq:perp-coh-len-from-ang-var} leads to an interesting interpretation of $\lcoh_\perp$.
It corresponds to the length in the transverse direction at which the incoherent sum of plane waves rotated by the angle $\tavg{\theta^2}^{1/2}$ are completely out of phase.
This is consistent with the intuitive definition of a coherence decay length as the characteristic length for the loss of phase relation.

\section{Conclusions}\label{sec:conclusions}
In this paper, some properties of the quantum master equation~\eqref{eq:simplified-redfield} governing the evolution of a fast particle in a gas, and derived in the previous paper~\cite{GaspardD2022c}, were studied in details.
The paper began in Sec.~\ref{sec:general-properties} with the presentation of the master equation and two of its major properties, namely thermalization and spatial decoherence.
\par Then, the master equation~\eqref{eq:simplified-redfield} was approximated in Sec.~\ref{sec:kramers-moyal} using the quantum counterpart of the Kramers-Moyal expansion to small momentum transfer.
This approximation is relevant if the particle is heavy compared to the bath scatterers or if the differential cross section is very peaked in the forward direction.
The Kramers-Moyal expansion led to the general form~\eqref{eq:caldeira-leggett-1} of the Caldeira-Leggett master equation valid for finite bath temperature~\cite{Caldeira1983b, Breuer2002, Diosi1993b, Vacchini2009, Hornberger2009, Kamleitner2010}.
This equation was then specialized to the case of strongly forward scattering, resulting in Eq.~\eqref{eq:caldeira-leggett-3}.
In particular, this equation contains a term of double commutator of the density matrix with the angular momentum operator, which is interpreted as the transverse decoherence term applying a random infinitesimal rotation to the particle wave packet.
In the Wigner representation, Eq.~\eqref{eq:caldeira-leggett-3} reduces to the Fokker-Planck equation~\eqref{eq:fokker-planck-fast}.
\par Furthermore, the coherence length of the particle was introduced and studied in Sec.~\ref{sec:coherence-length}.
The coherence length matrix was defined in Eq.~\eqref{eq:def-coherence-length-matrix} as the covariance matrix of the off-diagonal slice of the density matrix.
It was shown that, for a Gaussian state, the coherence length matrix is proportional to the inverse of the momentum covariance matrix.
This property can be interpreted as a consequence of the Heisenberg uncertainty principle.
In addition, it led, through the momentum variances of Eq.~\eqref{eq:def-momentum-variances}, to the time evolution of the coherence lengths in the directions parallel and perpendicular to the particle motion assuming that the transport parameters are independent of the particle energy.
Since the momentum spreads more quickly in the transverse direction than in the longitudinal direction, the coherence length is smaller in the transverse direction than in the longitudinal one.
Therefore, the coherent wave packet is more elongated in the direction of motion.
At short time, the ratio of both is given by Eq.~\eqref{eq:coherence-lengths-ratio}, which only depends on the velocities of the particle and the scatterers.
Moreover, it turns out that the transverse coherence length drops below the thermal wavelength before reaching it after a sufficiently long time.
\par The original question asked by Darwin and Mott~\cite{Darwin1929, Mott1929} was about the emergence of classical phenomena from plain quantum mechanics, such as the appearance of linear tracks of alpha particles in cloud chambers.
Another important question intimately related to the previous one is the nature of the quantum state of a particle propagating in a gaseous detector.
In this paper, this question was addressed within the theory of open quantum systems using the formalism of the reduced density matrix.
The evolution of the density matrix is governed by quantum master equations, such as the Redfield equation~\eqref{eq:simplified-redfield}, which reproduces many of the classical phenomena regarding the propagation of the particle, including ballistic transport, diffusion, and thermalization.
In addition, these equations are able to describe fundamental quantum phenomena, and especially spatial decoherence, which is believed to play a key role in the description of measurement in quantum mechanics~\cite{Zeh1970, Zeh1973, Joos1985, Zurek1991}.
Decoherence manifests itself as the evanescence in time of the off-diagonal elements of the reduced density matrix.
In this paper, it was shown that the decoherence of the particle in a gas occurs mainly in position space.
Furthermore, when the cross section peaks in the forward direction, decoherence is stronger in the transverse direction than in the longitudinal direction.
Therefore, the coherent wave packet should look like an ellipsoid elongated in the direction of motion.
This result is quite different from the spherical coherent wave packets predicted by the Caldeira-Leggett equation, which only applies to slow Brownian particles relatively close to equilibrium with the gas.
In contrast, the present paper accounts for the angular variation of the cross section, which cannot be neglected when the particle is much faster than the particles of the thermal bath.
This angular variation is, in particular, responsible for the non-trivial shape of the coherent wave packet that was already discussed above.
Finally, by highlighting anisotropies in spatial decoherence, the present work achieves a significant advance in the characterization of the state of a particle undergoing quantum measurement in a detector.
\par Last but not least, an important issue that should be investigated in future works concerns the treatment of the Coulomb interaction between the alpha particle and the gas.
When deriving the Boltzmann equation in the previous paper~\cite{GaspardD2022c}, it was assumed that the range of the interaction between the particle and the scatterers is short, but this is not the case for the Coulomb interaction, which is typically long ranged.
Nevertheless, the Coulomb interaction is most of the time screened by the electric charges in the medium.
In neutral molecules, the screening length can be of the order of the atomic Bohr radius, but, in the present context of fast particles, this length turns out to significantly exceed the atomic Bohr radius~\cite{Sigmund2006, Sigmund2014, Lifschitz1998}.
This is due to the fact that, when the particle velocity is greater than the electron velocities in the medium, the electrons do not have time to completely screen the electric field of the particle, resulting in long-range electric interactions between the particle and the molecules.
As a consequence, the particle may interact with several molecules at a time, and collective effects may occur in the medium~\cite[chap.~5]{Sigmund2006}.
The treatment of these effects typically resorts to dielectric linear response theory \cite{Tamm1991, Fermi1939, Fermi1940, Bohr1948, Halpern1948, Sigmund2006}.
The point is that these collective effects could influence the results obtained in this paper and the previous one~\cite{GaspardD2022c}, in particular through modifications of the master equations, which suggests interesting new directions of research.

\begin{acknowledgments}%
The authors are grateful to Prof.\ Pierre Gaspard for useful suggestions and for reviewing this manuscript.
The authors also thank Alban Dietrich for bringing Ref.~\cite{Lifschitz1998} to their attention.
This work was funded by the Belgian National Fund for Scientific Research (F.R.S.-FNRS) as part of the ``Research Fellow'' (ASP - Aspirant) fellowship program.
\end{acknowledgments}%

\appendix%
\section{Decoherence on a hypersphere}\label{app:decoherence-on-sphere}
The purpose of this appendix is to prove Eqs.~\eqref{eq:angular-momentum-property} and~\eqref{eq:wigner-spherical-decoherence} for the angular momentum operator $\op{L}_{ij}$ under the approximation~\eqref{eq:diagonality-approx}.
Using the definition~\eqref{eq:def-angular-momentum} of $\op{L}_{ij}$, one writes
\begin{equation}\label{eq:angular-momentum-demo-1}
\frac{1}{2}[\op{L}_{ij}, [\op{L}_{ij}, \op{\rho}_\sys]] = \underbrace{[\op{r}_i\op{k}_{\sys,j}, [\op{r}_i\op{k}_{\sys,j}, \op{\rho}_\sys]]}_{\op{C}_1} - \underbrace{[\op{r}_i\op{k}_{\sys,j}, [\op{r}_j\op{k}_{\sys,i}, \op{\rho}_\sys]]}_{\op{C}_2}  \:,
\end{equation}
where the implicit summation of repeated indices is used.
Note the order of the indices $i$ and $j$ on the right-hand side of Eq.~\eqref{eq:angular-momentum-demo-1}.
This double commutator of $\op{\rho}_\sys$ with $\op{L}_{ij}$ represents the transverse decoherence of the particle due to the change of direction caused by the collisions.
First, let us consider the term called $\op{C}_1$. Expanding the nested commutators leads to
\begin{equation}\label{eq:angular-momentum-c1-step-1}\begin{split}
\op{C}_1 & = [\op{r}_i\op{k}_{\sys,j}, [\op{r}_i\op{k}_{\sys,j}, \op{\rho}_\sys]]  \\
 & = \op{r}_i\op{k}_{\sys,j}\op{r}_i\op{k}_{\sys,j}\op{\rho}_\sys - 2\op{r}_i\op{k}_{\sys,j}\op{\rho}_\sys\op{r}_i\op{k}_{\sys,j} + \op{\rho}_\sys\op{r}_i\op{k}_{\sys,j}\op{r}_i\op{k}_{\sys,j}  \:.
\end{split}\end{equation}
Since $\op{\rho}_\sys$ is assumed to be quasi-diagonal in the momentum basis according to Eq.~\eqref{eq:diagonality-approx}, it is appropriate to move the wavevector components closer to $\op{\rho}_\sys$ using the canonical commutation relation
\begin{equation}\label{eq:canonical-commutation}
[\op{r}_i, \op{k}_{\sys,j}] = \I\delta_{ij}  \:.
\end{equation}
Applying this idea to the first two terms of Eq.~\eqref{eq:angular-momentum-c1-step-1} yields
\begin{equation}\label{eq:angular-momentum-c1-step-2}
\op{C}_1 = \op{r}_i\op{r}_i\op{\vect{k}}_\sys^2\op{\rho}_\sys - 2\op{r}_i\op{\vect{k}}_\sys^2\op{\rho}_\sys\op{r}_i - 3\I\op{r}_i\op{k}_{\sys,i}\op{\rho}_\sys + \op{\rho}_\sys\op{r}_i\op{k}_{\sys,j}\op{r}_i\op{k}_{\sys,j}  \:.
\end{equation}
The last term of Eq.~\eqref{eq:angular-momentum-c1-step-2} needs three commutations of the position and the momentum. The result is
\begin{equation}\label{eq:angular-momentum-c1-last}\begin{split}
\op{\rho}_\sys\op{r}_i\op{k}_{\sys,j}\op{r}_i\op{k}_{\sys,j} & = \op{\rho}_\sys(\op{k}_{\sys,j}\op{r}_i + \I\delta_{ij})(\op{k}_{\sys,j}\op{r}_i + \I\delta_{ij})  \\
 & = \op{\rho}_\sys\op{k}_{\sys,j}\op{r}_i\op{k}_{\sys,j}\op{r}_i + 2\I\op{\rho}_\sys\op{k}_{\sys,i}\op{r}_i - \op{\rho}_\sys d  \\
 & = \op{\rho}_\sys\op{\vect{k}}_\sys^2\op{r}_i\op{r}_i + 3\I\op{\rho}_\sys\op{k}_{\sys,i}\op{r}_i - \op{\rho}_\sys d  \:.
\end{split}\end{equation}
Then, substituting the result~\eqref{eq:angular-momentum-c1-last} back into Eq.~\eqref{eq:angular-momentum-c1-step-2} gives
\begin{equation}\label{eq:angular-momentum-c1-final}
\op{C}_1 = [\op{r}_i, [\op{r}_i, \op{\vect{k}}_\sys^2\op{\rho}_\sys]] - 3\I[\op{r}_i, \op{k}_{\sys,i}\op{\rho}_\sys] - \op{\rho}_\sys d  \:.
\end{equation}
Note that between Eqs.~\eqref{eq:angular-momentum-c1-step-2} and~\eqref{eq:angular-momentum-c1-final}, the assumption $[\op{\vect{k}}_\sys, \op{\rho}_\sys]\simeq 0$ has been used.
The term called $\op{C}_2$ in Eq.~\eqref{eq:angular-momentum-demo-1} can be calculated in the same way, but leads to a markedly different result.
Expanding the double commutator leads to
\begin{equation}\label{eq:angular-momentum-c2-step-1}\begin{split}
\op{C}_2 & = [\op{r}_i\op{k}_{\sys,j}, [\op{r}_j\op{k}_{\sys,i}, \op{\rho}_\sys]]  \\
 & = \op{r}_i\op{k}_{\sys,j}\op{r}_j\op{k}_{\sys,i}\op{\rho}_\sys - 2\op{r}_i\op{k}_{\sys,j}\op{\rho}_\sys\op{r}_j\op{k}_{\sys,i} + \op{\rho}_\sys\op{r}_i\op{k}_{\sys,j}\op{r}_j\op{k}_{\sys,i}  \:.
\end{split}\end{equation}
As before, one commutes the positions and momenta in order to get the momenta closer to $\op{\rho}_\sys$:
\begin{equation}\label{eq:angular-momentum-c2-step-2}\begin{split}
\op{C}_2 & = \op{r}_i\op{r}_j\op{k}_{\sys,i}\op{k}_{\sys,j}\op{\rho}_\sys - \I(d+2)\op{r}_i\op{k}_{\sys,i}\op{\rho}_\sys \\
 & - 2\op{r}_i\op{k}_{\sys,i}\op{k}_{\sys,j}\op{\rho}_\sys\op{r}_j + \op{\rho}_\sys\op{r}_i\op{k}_{\sys,j}\op{r}_j\op{k}_{\sys,i}  \:.
\end{split}\end{equation}
The last term of Eq.~\eqref{eq:angular-momentum-c1-step-2} needs also three commutation steps. They read
\begin{equation}\label{eq:angular-momentum-c2-last}\begin{split}
\op{\rho}_\sys\op{r}_i\op{k}_{\sys,j}\op{r}_j\op{k}_{\sys,i} & = \op{\rho}_\sys(\op{k}_{\sys,j}\op{r}_i + \I\delta_{ij})(\op{k}_{\sys,i}\op{r}_j + \I\delta_{ij})  \\
 & = \op{\rho}_\sys\op{k}_{\sys,j}\op{r}_i\op{k}_{\sys,i}\op{r}_j + 2\I\op{\rho}_\sys\op{k}_{\sys,i}\op{r}_i - \op{\rho}_\sys d  \\
 & = \op{\rho}_\sys\op{k}_{\sys,i}\op{k}_{\sys,j}\op{r}_i\op{r}_j + \I(d+2)\op{\rho}_\sys\op{k}_{\sys,i}\op{r}_i - \op{\rho}_\sys d  \:.
\end{split}\end{equation}
Inserting Eq.~\eqref{eq:angular-momentum-c2-last} back into Eq.~\eqref{eq:angular-momentum-c2-step-2} leads to the result
\begin{equation}\label{eq:angular-momentum-c2-final}
\op{C}_2 = [\op{r}_i, [\op{r}_j, \op{k}_{\sys,i}\op{k}_{\sys,j}\op{\rho}_\sys]] - \I(d+2) [\op{r}_i, \op{k}_{\sys,i}\op{\rho}_\sys] - \op{\rho}_\sys d  \:.
\end{equation}
Finally, combining Eqs.~\eqref{eq:angular-momentum-c1-final} and~\eqref{eq:angular-momentum-c2-final} into Eq.~\eqref{eq:angular-momentum-demo-1}, one obtains the sought property
\begin{equation}\label{eq:angular-momentum-demo-final}\begin{split}
\frac{1}{2}[\op{L}_{ij}, [\op{L}_{ij}, \op{\rho}_\sys]] & = [\op{r}_i, [\op{r}_j, (\op{\vect{k}}_\sys^2\delta_{ij} - \op{k}_{\sys,i}\op{k}_{\sys,j})\op{\rho}_\sys]]  \\
 & + \I(d-1) [\op{r}_i, \op{k}_{\sys,i}\op{\rho}_\sys]  \:,
\end{split}\end{equation}
which is used in Eq.~\eqref{eq:angular-momentum-property}.
\par One still has to determine the effect of the double commutator of $\op{\rho}_\sys$ with $\op{L}_{ij}$ in the Wigner representation, since this representation is used in Sec.~\ref{sec:wigner-representation}.
Because of the quasi-diagonality of $\op{\rho}_\sys$ in the momentum basis, it is related to the Wigner function $f_\sys(\vect{k}_\sys)$ by
\begin{equation}\label{eq:inverse-wigner}
\op{\rho}_\sys \simeq (2\pi)^d \int_{\mathbb{R}^d} \D\vect{k}_\sys f_\sys(\vect{k}_\sys) \ket{\vect{k}_\sys} \bra{\vect{k}_\sys}  \:.
\end{equation}
An important consequence is that the commutators with $\op{L}_{ij}$ reduce to a multiplication by $\op{L}_{ij}$ in the Wigner representation:
\begin{equation}\label{eq:angular-momentum-wigner-1}
\wigner{\frac{1}{2}[\op{L}_{ij}, [\op{L}_{ij}, \op{\rho}_\sys]]} = \frac{1}{2}\op{L}_{ij}\op{L}_{ij} f_\sys  \:.
\end{equation}
It should be noted that on the right-hand side of Eq.\ \eqref{eq:angular-momentum-wigner-1}, the two operators $\op{L}_{ij}=\op{r}_i\op{k}_{\sys,j} - \op{r}_j\op{k}_{\sys,i}$ implicitly assume that
\begin{equation}\label{eq:operator-analogy}
\op{r}_i = \I\pder{}{k_{\sys,i}}  \qquad\text{and}\qquad  \op{k}_{\sys,i} = k_{\sys,i}  \:.
\end{equation}
Of course, these equalities follow from the principle of correspondence in quantum mechanics.
They are expressed in the momentum basis, and not in the position basis, because of the quasi-diagonality of $\op{\rho}_\sys$ assumed in Eq.~\eqref{eq:inverse-wigner}.
The scalar product of angular momenta can now be expanded from the definition~\eqref{eq:def-angular-momentum}.
Using the canonical commutation relation, it is relatively straightforward to get
\begin{equation}\label{eq:angular-momentum-expansion-1}
\frac{1}{2}\op{L}_{ij}\op{L}_{ij} = \op{\vect{k}}_\sys^2\op{\vect{r}}^2 - (\op{\vect{k}}_\sys\cdot\op{\vect{r}})^2 - \I(d-2) \op{\vect{k}}_\sys\cdot\op{\vect{r}}  \:.
\end{equation}
The equalities in Eq.~\eqref{eq:operator-analogy} directly lead to
\begin{equation}\label{eq:angular-momentum-expansion-2}
\frac{1}{2}\op{L}_{ij}\op{L}_{ij} = -k_\sys^2\lapl_{\vect{k}_\sys} + (\vect{k}_\sys\cdot\grad_{\vect{k}_\sys})^2 + (d-2) \vect{k}_\sys\cdot\grad_{\vect{k}_\sys}  \:.
\end{equation}
Using the fact that $\vect{k}_\sys\cdot\grad_{\vect{k}_\sys}=k_\sys\partial_{k_\sys}$, one gets
\begin{equation}\label{eq:angular-momentum-expansion-3}
\frac{1}{2}\op{L}_{ij}\op{L}_{ij} = -k_\sys^2\lapl_{\vect{k}_\sys} + k_\sys^2\pder[2]{}{k_\sys} + (d-1) k_\sys\pder{}{k_\sys}  \:,
\end{equation}
The last step is to expand the Laplacian operator in spherical coordinates~\cite{Olver2010}
\begin{equation}\label{eq:laplacian-spherical-coords}
\lapl_{\vect{k}_\sys} = \pder[2]{}{k_\sys} + \frac{d-1}{k_\sys}\pder{}{k_\sys} + \frac{1}{k_\sys^2}\lapl_{\perp\vect{k}_\sys}  \:,
\end{equation}
where $\lapl_{\perp\vect{k}_\sys}$ denotes the spherical Laplacian, which acts only on the unit hypersphere in the space $\mathbb{R}^d$.
Substituting Eq.~\eqref{eq:laplacian-spherical-coords} into Eq.~\eqref{eq:angular-momentum-expansion-3} yields
\begin{equation}\label{eq:angular-momentum-expansion-final}
\frac{1}{2}\op{L}_{ij}\op{L}_{ij} = -\lapl_{\perp\vect{k}_\sys}  \:.
\end{equation}
This result shows that the action of the double commutator in Eq.~\eqref{eq:angular-momentum-wigner-1} corresponds to a spherical Laplacian in the momentum space.
This also proves Eq.~\eqref{eq:wigner-spherical-decoherence}.

\bibliographystyle{apsrev4-2-fixed}%
\end{document}